\documentclass[11pt, draftclsnofoot, onecolumn]{IEEEtran}
\usepackage{cite}
\usepackage{balance}

\ifCLASSINFOpdf
\else
\fi

\usepackage{tcolorbox}
\usepackage{amsmath,amssymb,epsfig,psfrag,cite,subfigure}
\usepackage{amsfonts,bm}
\usepackage{graphicx}
\usepackage{epstopdf}

\usepackage{hyperref}
\usepackage{microtype}
\usepackage{setspace}
\usepackage{tcolorbox}
\newcommand{\rev}[1]{{#1}}

\usepackage[nolist]{acronym}
\begin{acronym}[ACRONYM]
\acro{3GPP}{The 3rd Generation Partnership Project}
\acro{3D}{three dimensional}
\acro{5G}{fifth generation}
\acro{6G}{sixth generation}
\acro{AI}{artificial intelligence}
\acro{AR}{augmented reality}
\acro{AoA}{angle-of-arrival}
\acro{AoD}{angle-of-departure}
\acro{BS}{base station}
\acro{CF}{consumption factor}
\acro{CSI}{channel state information}
\acro{DL}{downlink}
\acro{D-MIMO}{distributed multiple-input multiple-output}
\acro{EER}{energy efficiency ratio}
\acro{EM}{electromagnetic}
\acro{GDOP}{geometric dilution of precision}
\acro{GNSS}{global navigation satellite system}
\acro{HMI}{human-machine interaction}
\acro{IQ}{in-phase and quadrature}
\acro{JCS}{Joint Communication and Sensing}
\acro{ISAC}{integrated sensing and communication}
\acro{JRC}{joint radar and communication}
\acro{JRC2LS}{joint radar communication, computation, localization, and sensing}
\acro{ICI}{inter-carrier interference}
\acro{IOO}{indoor open office}
\acro{IoT}{Internet of Things}
\acro{IRN}{infrastructure reference node}
\acro{ISG}{industry specification group}
\acro{JCAS}{joint communications and sensing}
\acro{KPI}{key performance indicator}
\acro{KVI}{key value indicator}
\acro{LCA}{life cycle analysis}
\acrodefplural{LCA}{life cycle analyses}
\acro{LoS}{line-of-sight}
\acro{MAC}{medium access control}
\acro{MIMO}{multiple-input multiple-output}
\acro{ML}{machine learning}
\acro{mmWave}{millimeter-wave}
\acro{NF}{network function}
\acro{NLoS}{non-line-of-sight}
\acro{NR}{new radio}
\acro{NTN}{non-terrestrial network}
\acro{OFDM}{orthogonal frequency-division multiplexing}
\acro{OTFS}{orthogonal time-frequency-space}
\acro{PRS}{positioning reference signal}
\acro{PHY}{physical layer}
\acro{QoS}{Quality of Service}
\acro{RAN}{radio access network}
\acro{RAT}{radio access technology}
\acro{RedCap}{reduced capacity}
\acro{RF}{radio frequency}
\acro{RIS}{reconfigurable intelligent surface}
\acro{RTK}{real-time kinematic}
\acro{RTT}{round-trip-time}
\acro{SDG}{sustainable development goal} 
\acro{SLAM}{simultaneous localization and mapping}
\acro{SNR}{signal-to-noise ratio}
\acro{SIT}{sustainability, inclusiveness, and trustworthiness}
\acro{SOTA}{state of the art}
\acro{SL}{sidelink}
\acro{ToA}{time-of-arrival}
\acro{TDoA}{time-difference-of-arrival}
\acro{TR}{time-reversal}
\acro{TRP}{transmission and reception point}
\acro{TXRX}[TX/RX]{transmitter/receiver}
\acro{TX}{transmitter}
\acro{RX}{receiver}
\acro{UE}{user equipment}
\acro{UN}{United Nations}
\acro{multi-RTT}{multi-cell round-trip-time}
\acro{UL}{uplink}
\acro{UL-TDOA}{uplink time-difference-of-arrival}
\acro{DL-TDOA}{downlink time-difference-of-arrival}
\acro{UMi}{3D-urban micro}
\acro{UMa}{3D-urban macro}
\acro{UWB}{ultra-wide band}
\acro{FR1}{frequency range 1}
\acro{FR2}{frequency range 2}
\acro{WLAN}{wireless local-area network }

\end{acronym}
\begin{document}
\newcommand{\txrm}{{{\rm{T}}}}
\newcommand{\rxrm}{{{\rm{R}}}}
\newcommand{\comrm}{{{\rm{com}}}}
\newcommand{\radrm}{{{\rm{rad}}}}
\newcommand{\thn}[1]{ {#1^{\rm{th} } } }

\newcommand{\complexset}[2]{ \mathbb{C}^{#1 \times #2}  }
\newcommand{\complexsett}{ \mathbb{C}  }
\newcommand{\realset}[2]{ \mathbb{R}^{#1 \times #2}  }
\newcommand{\nset}{ \mathbb{N}  }
\newcommand{\zset}{ \mathbb{Z}  }

\newcommand{\trp}{^\top}
\newcommand{\herm}{^\mathrm{H}}

\newcommand{\boldzero}{{ {\boldsymbol{0}} }}
\newcommand{\boldone}{{ {\boldsymbol{1}} }}
\newcommand{\boldonet}{{ {\boldsymbol{1}}^{T} }}
\newcommand{\Imatrix}{{ \boldsymbol{\mathrm{I}} }}

\newcommand{\realp}[1]{ \Re \left\{#1\right\}  }
\newcommand{\imp}[1]{ \Im \left\{#1\right\}  }
\newcommand{\vecc}[1]{ {\rm{vec}}\left(#1\right)  }
\newcommand{\veccs}[1]{ {\rm{vec}}\big(#1\big)  }
\newcommand{\veccb}[1]{ {\rm{vec}}\Big(#1\Big)  }

\newcommand{\aaa}{\mathbf{a}}
\newcommand{\cc}{ \mathbf{c} }
\newcommand{\bb}{ \mathbf{b} }
\newcommand{\nn}{ \mathbf{n} }
\newcommand{\hh}{ \mathbf{h} }
\newcommand{\hhfreqc}{\hh_{\rm{c,freq}}}

\newcommand{\ffb}{ \mathbf{f} }
\newcommand{\ww}{ \mathbf{w} }
\newcommand{\vv}{ \mathbf{v} }

\newcommand{\Tsym}{ T_{\rm{sym}} }
\newcommand{\Tcp}{ T_{\rm{cp}} }
\newcommand{\Ttot}{ T_{\rm{tot}} }

\newcommand{\cfo}{ \delta_f }
\newcommand{\bias}{ \delta_\tau }
\newcommand{\cfoc}{ \delta'_f }

\newcommand{\deltaf}{ \Delta f }
\newcommand{\deltafc}{ \Delta \fc }
\newcommand{\fc}{ f_c }
\newcommand{\Ntx}{ N_\txrm }
\newcommand{\Nrx}{ N_\rxrm }
\newcommand{\atx}{ \aaa_{\rm{t}} }
\newcommand{\arx}{ \aaa_{\rm{r}} }
\newcommand{\arxc}{ \aaa_{\rm{c}} }

% transmit, sensing RX and comm RX antenna elements
\newcommand{\Nt}{ N_{\rm{t}} }
\newcommand{\Nss}{ N_{\rm{r}}  }
\newcommand{\Nc}{ N_{\rm{c}}  }

\newcommand{\ynm}{ y_{n,m} }
\newcommand{\xnm}{ x_{n,m} }
\newcommand{\pnm}{ P_{n,m} }

\newcommand{\XX}{ \mathbf{X} }
\newcommand{\PP}{ \mathbf{P} }
\newcommand{\RR}{ \mathbf{R} }
\newcommand{\WW}{ \mathbf{W} }
\newcommand{\VV}{ \mathbf{V} }
\newcommand{\DD}{ \mathbf{D} }
\newcommand{\YY}{ \mathbf{Y} }
\newcommand{\ZZ}{ \mathbf{Z} }
\newcommand{\JJ}{ \mathbf{J} }
\newcommand{\HH}{ \mathbf{H} }
\newcommand{\FF}{ \mathbf{F} }

\newcommand{\YYr}{ \YY_{\rm{r}} }
\newcommand{\YYc}{ \YY_{\rm{c}} }

\newcommand{\HHc}{ \HH_{\rm{c}} }
\newcommand{\ZZc}{ \ZZ_{\rm{c}} }
\newcommand{\ZZr}{ \ZZ_{\rm{r}} }
\newcommand{\sigmac}{ \sigma_{\rm{c}} }
\newcommand{\sigmar}{ \sigma_{\rm{r}} }

\newcommand{\Ns}{N_s}
\newcommand{\Lt}{L_\txrm}
\newcommand{\Lr}{L_\rxrm}
\newcommand{\bF}{\mathbf{F}}
\newcommand{\Frf}{\bF_{\rm{RF}}}
\newcommand{\Fbb}{\bF_{\rm{BB}}}
\newcommand{\bW}{\mathbf{W}}
\newcommand{\Wrf}{\bW_{\rm{RF}}}
\newcommand{\Wbb}{\bW_{\rm{BB}}}

\newcommand{\bz}{\mathbf{z}}
\newcommand{\by}{\mathbf{y}}
\newcommand{\bx}{\mathbf{x}}
\newcommand{\tildex}{\tilde{\bx}}
\newcommand{\bH}{\mathbf{H}}

\newcommand{\taubar}{\bar{\tau}}
\newcommand{\nubar}{\bar{\nu}}

\newcommand{\AT}{\mathbf{A}_\txrm }
\newcommand{\AR}{\mathbf{A}_\rxrm }
\newcommand{\btheta}{\boldsymbol \theta}
\newcommand{\bphy}{\boldsymbol \phy}
\newcommand{\Hrnm}{\mathbf{H}_{\text{r},n,m}}

\newcommand{\pd}{ P_{\rm{d}} }
\newcommand{\pfa}{ P_{\rm{fa}} }

\newcommand{\etabhat}{ \widehat{\boldsymbol{\eta}} }
\newcommand{\etab}{ {\boldsymbol{\eta}} }

\newcommand{\sml}{ s_{m,\ell} }
\newcommand{\sm}{ s_{m} }

\newcommand{\rect}[1]{ { \rm{rect} }\left(#1\right) }

\newcommand{\mtCN}{{\mathcal{CN}}}
\newcommand{\ptot}{P_{\mathrm{tot}}}

\newcommand{\Hzero}{\mathcal{H}_0}
\newcommand{\Hone}{\mathcal{H}_1}

\newcommand{\norm}[1]{\left\lVert#1\right\rVert}
\newcommand{\normsmall}[1]{\big\lVert#1\big\rVert}
\newcommand{\normbig}[1]{\Big\lVert#1\Big\rVert}

\newcommand{\normsq}[1]{ \norm{#1}^{2} }
\newcommand{\normf}[1]{ \norm{#1}_{\rm{F}}^{2} }
\newcommand{\normfsmall}[1]{ \normsmall{#1}_{\rm{F}}^{2} }

\newcommand{\Eee}{\mathbb{E}}
\newcommand{\quot}[1]{``{#1}''}

\newcommand{\diag}[1]{ {\rm{diag}}\left(#1\right)  }
\newcommand{\ddiag}[1]{ {\rm{ddiag}}\left(#1\right)  }
\newcommand{\blkdiag}[1]{ {\rm{blkdiag}}\left(#1\right)  }
\newcommand{\ncoll}[1]{N_{\text{C}}\left[ #1 \right] }
\newcommand{\nmiss}[1]{N_{\text{M}}\left[ #1 \right] }

\newcommand{\cqun}{q_{u,n}}
\newcommand{\cqsn}{q_{s,n}}
\newcommand{\cqin}{q_{i,n}}
\newcommand{\cqsnl}{q_{s,n}^{(l)}}
\newcommand{\cqinl}{q_{i,n}^{(l)}}

\newcommand{\snr}{{\rm{SNR}}}
\newcommand{\gammasnr}{\gamma_{\rm{SNR}}}

% Operators
\newcommand{\tr}[1]{{#1}^\mathrm{T}}
\renewcommand{\vec}[1]{\bm{\mathrm{#1}}}
\newcommand{\E}[1]{\mathbb{E}\left[ #1 \right]}
\newcommand{\Earg}[2]{\mathbb{E}_{#1}\left[ #2 \right]}
\newcommand{\var}[1]{\mathrm{Var}\left[{#1}\right]}
\newcommand{\svec}[1]{\bm{#1}}
\newcommand{\re}[1]{\text{Re}\left\{ #1 \right\}}
\newcommand{\pr}[1]{\text{Pr}\left\{#1\right\}}
\newcommand{\sigmoid}[1]{\sigma\left( #1 \right)}

% Special vectors
\newcommand{\ones}{\vec{1}}
\newcommand{\zeros}{\vec{0}}

\newcommand{\indic}[1]{{\ones}_{\{#1\}}}

\newcommand{\s}{ \mathbf{s} }
%\renewcommandx{\a}[2][1=default, 2=default]{\param{a}[#1][#2]} % Control
%\newcommand{\a}{ \mathbf{a} }
%\newcommandx{\z}[2][1=default, 2=default]{\param{z}[#1][#2]} % Observation
\newcommand{\z}{ \mathbf{z} }
%\newcommandx{\p}[2][1=default, 2=default]{\param{p}[#1][#2]} % A priori occupied probability
\newcommand{\p}{ \mathbf{p} }
%\%newcommandx{\y}[2][1=default, 2=default]{\param{\hat{s}}[#1][#2]} % A priori occupied probability
\newcommand{\y}{ \mathbf{y} }
%% Other

%\include{macros}
\bstctlcite{IEEEexample:BSTcontrol}

\title{Multicarrier ISAC: Advances in Waveform Design,
Signal Processing and Learning under Non-Idealities %{\textcolor{red}{20 pages, 30 refs}}
}

\author{Visa Koivunen, %~\IEEEmembership{Fellow,~IEEE,}
        Musa Furkan Keskin, %~\IEEEmembership{Member, IEEE,}
        Henk Wymeersch, %~\IEEEmembership{Fellow,~IEEE,}
        Mikko Valkama, \\and %~\IEEEmembership{Fellow,~IEEE,}
        Nuria González-Prelcic %~\IEEEmembership{Senior Member,~IEEE}% <-this % stops a space
\thanks{V. Koivunen is with Aalto University, Finland. M. F. Keskin and H. Wymeersch are with Chalmers University of Technology, Sweden. M. Valkama is with Tampere University, Finland. N. González-Prelcic is with University of California  San Diego, USA.} 

}

% make the title area
\maketitle

\vspace{-15mm}
\begin{abstract}

    This paper addresses the topic of integrated sensing and communications (ISAC) in 5G and emerging 6G wireless networks. ISAC systems operate within shared, congested or even contested spectrum, aiming to deliver high performance in both wireless communications and radio frequency (RF) sensing. The expected benefits include more efficient utilization of spectrum, power, hardware (HW) and antenna resources. Focusing on multicarrier (MC) systems, which represent the most widely used communication waveforms, it explores the co-design and optimization of waveforms alongside multiantenna transceiver signal processing for communications and both monostatic and bistatic sensing applications of ISAC. Moreover, techniques of high practical relevance for overcoming and even harnessing challenges posed by non-idealities in actual transceiver implementations are considered. To operate in highly dynamic radio environments and target scenarios, both model-based structured optimization and learning-based methodologies for ISAC systems are covered, assessing their adaptability and learning capabilities under real-world conditions. The paper presents trade-offs in communication-centric and radar-sensing-centric approaches, aiming for an optimized balance in densely used spectrum.

    \vspace{-5mm}
\end{abstract}
%\newpage 
\section{Introduction}

Integrated sensing and communications (ISAC) systems operate within shared, congested, or even contested spectrum, aiming to deliver high performance in both wireless communications and radio frequency (RF) sensing \cite{bliss17,Fan_ISAC_6G_JSAC_2022}. The expected benefits include more efficient utilization of spectrum, power, hardware (HW) and antenna resources. Sensing and communication systems cooperate and may even be co-designed for mutual benefit. Particularly in the context of ISAC for 6G networks, 
there is significant interest in multi-user communications and RF sensing systems that may share hardware and antenna resources, use joint waveforms, and function within shared spectra and exchange awareness about their radio environments to optimize performance \cite{zheng2019radar}. {ISAC technology has opened genuinely new lines of research and development rather than merely being an evolution of 5G systems. }

Accurate user localization and its extension to jointly estimating also the environment scattering points, or the so-called landmarks, in the spirit of simultaneous localization and mapping (SLAM) \cite{Ge2022JSAC} are timely examples.

This paper focuses on multicarrier (MC) ISAC systems, especially waveform design and optimization, signal processing, adaptation, and learning for their implementation in practical transceivers that face various non-idealities. The focus on MC is motivated by the fact that 
most current and emerging wireless communication and broadcast systems, including 4G, 5G and future 6G, WLAN/WiFi, 
 {digital video broadcasting–terrestrial (DVB-T/T2)} and  {digital audio broadcasting (DAB)}, employ MC waveforms that use a large number of orthogonal subcarriers. Orthogonal frequency-division multiplexing (OFDM) is the most widely used MC waveform, often accompanied by orthogonal frequency-division multiple access (OFDMA) for multiuser scenarios and multiantenna-based {multiple-input multiple-output (MIMO)} technologies. 
OFDM was adopted for wireless communication standards like WiFi and WiMAX, as well as  multiantenna OFDM(A)-based 4G systems. 
OFDM(A) continued into 5G, leveraging massive-MIMO and millimeter-wave frequencies. 

Multicarrier waveforms are widely used for radar tasks, for example, the multi-carrier phase coded (MCPC) waveforms \cite{levanon2000multifrequency}, OFDM-based radar \cite{Sturm11} and as signals of opportunity {in a variety of passive radars.} 

Modern commercial communications systems 
take advantage of highly efficient hardware implementations 
with rapid integrated circuit (IC) design cycles. Hence, MC transceiver design  can also exploit these fast design cycles, the cost and power efficient circuits, hardware and antenna system implementations \cite{bliss17,FD_LTE_NR_OFDM_2019} for ISAC. This is
particularly beneficial for radar system development since traditionally their HW design cycles have been very long and costly, and the life span of a radar system may expand over multiple decades. 
ISAC takes advantage of the ongoing parallel convergence of  
multi-function HW, RF circuitry, adaptive large aperture multi-antenna systems and fully adaptive multi-function radars \cite{bliss17,hassanien2019dual}.
In addition, MC waveforms possess desirable properties for broadband communications and RF sensing. 
They not only provide robust communication performance through simple equalization, but also enable high-accuracy and low-complexity radar processing (e.g., via 2-D fast Fourier transform, FFT \cite{Sturm11}). In radar tasks, MC waveforms offer efficient Doppler processing, frequency diversity and reduced time-on-target (see \cite{bicatrsp16} and references therein).
MC waveforms bring adaptability, learning and large design flexibility via allocating subcarriers and powers for different users or tasks \cite{bliss17,hassanien2019dual}. Coding, waveform diversity, interference management and sharing resources in time, frequency, and spatial domains  facilitate adaptation to rapidly varying ISAC performance requirements, dynamic radio and target environments, and enables flexibly tunable ISAC trade-offs in real-world operation. 
All in all, MC waveforms provide a promising foundation for ISAC. 
There are several recent overview and tutorial papers on ISAC and dual function radar-communications (DFRC) systems, see  \cite{Fan_ISAC_6G_JSAC_2022}, \cite{zheng2019radar}, \cite{hassanien2019dual},  \cite{mishra2019toward} and \cite{ma2020joint} and references therein.
However, key topics on MC ISAC, including multifunction hardware and corresponding impairments, waveform design in time-frequency-space domains, and transceiver adaptation and learning in different ISAC tasks are only treated partially. Consequently, the potential of MC multiantenna systems is not fully taken advantage of. 

\begin{figure}
    \centering
    \includegraphics[width=0.99\columnwidth]{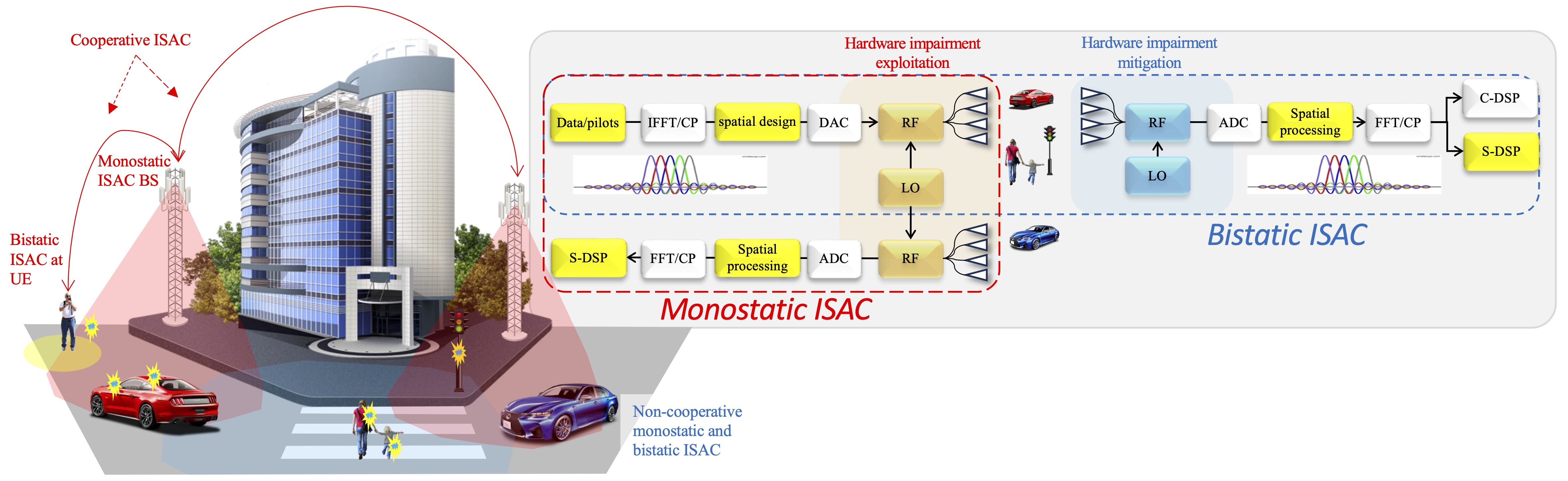}
    \vspace{-3mm}
    \caption{Overview of the scope of the paper. Cooperative and co-designed RF sensing and wireless communications subsystems use MC waveforms for mutual benefit while dealing with various impairments.}
    \vspace{-3mm}
    \label{fig:overview}
\end{figure}

This  article focuses on various signal processing  techniques employed to achieve end goals in a MC ISAC system and related optimization and adaptation of the MC waveforms, as illustrated in Fig.~\ref{fig:overview}. We start by detailing the models of MC waveforms from communication-centric and radar-centric perspectives along with the key performance indicators (KPI), while also describing channel models for monostatic and bistatic operation. Then, we describe the signal processing steps and the operation under hardware non-idealities (such as frequency offsets, phase noise (PN), power amplifier nonlinearities (PAN), IQ-imbalances, and array spacing perturbations \cite{FD_LTE_NR_OFDM_2019}), to extract the channel parameters, which are used for localization, mapping, and SLAM solutions. The last part of the paper deals with the co-design of sensing and communications and finding favorable ISAC trade-offs in  densely used radio spectrum \cite{pulkkinen24,end2end_ISAC_2022}. {The design, optimization and adaptation problems can be formulated as structured optimization problems or data-driven machine learning problems both at transmitters and receivers.}

%%%%%%%%%%%%%%%%%%%%%%%%%%%%%%%%%%%%%%%%%%%%%%%%%%%%%%%

\section{Models and Waveforms for Multicarrier ISAC Systems}

In this section, we provide an overview of commonly employed MC ISAC waveforms, including OFDM, MCPC and multicarrier code-division multiple access (MC-CDMA), as well as emerging MC waveforms towards 6G, along with receive signal models {and} KPIs. {In wireless communications systems and radars, multiantenna transceivers are typically used with multicarrier waveforms. However, in many broadcast systems such as DVB-T2 and DAB, single antenna transmitters are deployed, and the transmitted waveforms are  commonly used as signals of opportunity in passive radars \cite{Passive_OFDM_2010}}.

%%%%%%%%%%%%%%%%%%%%%%%%%%%%%%%%%%%%%%%%%%%%%%%%%%%%%%%
\subsection{Signal Models for {MIMO-}OFDM ISAC Systems} 
Consider a {MIMO-}OFDM ISAC system, comprising an {$\Nt$-antenna} ISAC transmitter (TX), a {$\Nss$-antenna} radar sensing receiver (RX) and a {$\Nc$-antenna} communications RX. In monostatic sensing, the ISAC TX is co-located with the radar RX on the same device and utilizes a shared oscillator. In bistatic sensing, they are located on separate devices in distinct locations and assumed to use independent oscillators. The ISAC TX transmits data/pilot symbols to the communications RX, while the sensing RX performs monostatic/bistatic radar sensing using the signals backscattered or reflected off the targets to the direction of RXs. \rev{We use the following notation: ${\mathbf a}$, ${\mathbf a}^*$, ${\mathbf a}\trp$ and ${\mathbf a}\herm$ for a column vector, its complex conjugate, transpose and Hermitian transpose, respectively; a matrix is denoted by ${\mathbf A}$, inner product of two vectors by ${\mathbf a}\trp {\mathbf b}$, and expectation of a random variable by $\Eee\{ X\}$.}

\subsubsection{Transmit Signal Model}
We consider an OFDM frame with $N$ subcarriers, $M$ symbols, subcarrier spacing $\deltaf$, elementary symbol duration $T = 1/\deltaf$, total symbol duration $\Tsym = \Tcp + T$ and the cyclic prefix (CP) duration $\Tcp$. Denoting by $\xnm$ {and $\pnm$} the complex data {and the transmit power, respectively,} on the $\thn{n}$ subcarrier of the $\thn{m}$ symbol, the complex baseband OFDM signal transmitted by the ISAC TX {for the $\thn{m}$ symbol} can be written as \cite{bicatrsp16,Sturm11,Fan_ISAC_6G_JSAC_2022}
\begin{align} \label{eq_st}
    {s_m(t)} = \frac{1}{\sqrt{N}}  \sum_{n = 0}^{N-1} {\sqrt{\pnm}}  \xnm \, e^{j 2 \pi n \deltaf t} \rect{\frac{t - m\Tsym}{\Tsym}}  ~,
\end{align}
where $\rect{t}$ is a rectangular pulse that takes the value $1$ for $t \in \left[0, 1 \right]$ and $0$ otherwise. {For power normalization, we set $\Eee\{ \lvert \xnm \rvert^2 \} = 1$ and $\sum_{m,n} \pnm = \ptot$, which denotes the total power constraint. \rev{For multi-antenna ISAC transmission, we consider a single-stream beamforming model \cite{80211_Radar_TVT_2018,MIMO_OFDM_ICI_JSTSP_2021}. Then,} given the baseband signal in \eqref{eq_st}, the upconverted \rev{multi-antenna transmit} signal over the entire frame is given by $\Re \left\{ \sum_{m=0}^{M-1} \ffb_m s_m(t) e^{j 2 \pi \fc t} \right\}$, where $\ffb_m \in \complexset{\Nt}{1}$ is the TX beamforming vector for the $\thn{m}$ symbol and $\fc$ denotes the carrier frequency. \rev{The generalization of this model to multiple stream  transmission and reception based on a hybrid beamforming architecture is described later in Sec.~\ref{sec_si} for the specific case of joint monostatic sensing and communication.}} 

\subsubsection{Radar Receive Signal Model}
We assume the presence of $K$ point targets in the environment, each characterized by an initial delay $\tau_k$, a Doppler shift $\nu_k$, {angle-of-departure (AoD) $\theta_k$, angle-of-arrival (AoA) $\phi_k$} and a complex channel gain $\alpha_k$, which involves scattering (radar cross section (RCS)), path loss and antenna gain effects. 
Given the transmit signal \eqref{eq_st}, the noiseless backscattered signal {for the $\thn{m}$ symbol} at the {$\thn{i}$ element of the} radar RX {array} after downconversion can be expressed as 
\begin{align}\label{eq_rec_baseband}
    {y_m^{i}(t)} &= \sum_{k=0}^{K-1} \alpha_k \, {[\arx(\phi_k)]_i \, \atx\trp(\theta_k) \ffb_m} {s_m(t-\tau_k(t))} e^{-j 2 \pi \fc \tau_k(t)} e^{j 2 \pi \cfo t}   ~,
\end{align}
where {$\atx(\theta) \in \complexset{\Nt}{1}$ and $\arx(\phi) \in \complexset{\Nss}{1}$ denote the array steering vectors at the ISAC TX and the radar RX, respectively,} $\tau_k(t) = \tau_k - \nu_k t /\fc + \bias$ is the time-varying delay due to target mobility, and $\cfo$ and $\bias$ are the carrier frequency offset (CFO) and clock offset between the ISAC TX and the {radar} RX, respectively. Note that in monostatic sensing $\cfo = 0$ and $\bias = 0$  due to the use of a shared oscillator, {and $\theta_k = \phi_k$ due to co-located TX and RX arrays}, while $\cfo \neq 0$, $\bias \neq 0$, {and $\theta_k \neq \phi_k$} in bistatic configurations, {where TX and RX arrays non-co-located}.  
In \eqref{eq_rec_baseband}, we set the clock reference (i.e., $t=0$) of the radar RX to the arrival time of the closest target echo (which can be detected, for example, via synchronization signals inserted at the beginning of 5G NR OFDM frames).
Plugging $\tau_k(t)$ and \eqref{eq_st} into ${y_m^{i}(t)}$ in \eqref{eq_rec_baseband}, and sampling ${y_m^{i}(t)}$ at $t = m\Tsym + \Tcp + \ell T / N$ for $\ell = 0, \ldots, N-1$ and $m = 0, \ldots, M-1$ (i.e., removing the CP for each symbol) yields \cite{Fan_ISAC_6G_JSAC_2022,PN_Exploitation_TSP_2023,MIMO_OFDM_ICI_JSTSP_2021}
\begin{align}\label{eq_rec_bb2}
    {y^{i}_{\ell,m}} &= \frac{1}{\sqrt{N}}  \sum_{k=0}^{K-1}  \sum_{n = 0}^{N-1} {\sqrt{\pnm}}  \xnm {\alpha_k^{i} \,\atx\trp(\theta_k) \ffb_m}  \, e^{-j 2 \pi n \deltaf \taubar_k} e^{j 2 \pi n \frac{\ell}{N}} e^{j 2 \pi T \frac{\ell}{N} \nubar_k } e^{j 2 \pi m \Tsym \nubar_k  }   ~,
\end{align}
which represents the signal at the $\thn{\ell}$ sample of the $\thn{m}$ symbol {at the $\thn{i}$ RX element, where $\alpha_k^{i} = \alpha_k [\arx(\phi_k)]_i$}. Following the radar nomenclature, the sample and symbol domains are denoted as \textit{fast-time} and \textit{slow-time}, respectively. In \eqref{eq_rec_bb2}, $\taubar_k = \tau_k + \bias$ and $\nubar_k = \nu_k + \cfo$. In obtaining \eqref{eq_rec_bb2} from \eqref{eq_rec_baseband}, we rely on the following standard assumption for OFDM sensing: cyclic prefix (CP) is larger than \textit{(i)} the delay spread of the targets in bistatic sensing ($\Tcp \geq \max_k \tau_k - \min_k \tau_k$) and \textit{(ii)} the round-trip delay of the furthermost target in monostatic sensing ($\Tcp \geq \max_k \tau_k$){\footnote{{Using a standard 5G NR numerology with $\deltaf = 30 \, \rm{kHz}$ \cite[Sec.~4.2]{TR_38211}, this assumption results in $\Tcp = 0.07/\deltaf = 2.33 \, \rm{\mu s}$. This translates to a maximum distance of $350 \, \rm{m}$ in monostatic sensing and a maximum distance spread of $700 \, \rm{m}$ in bistatic sensing. Such parameters are sufficient to address a wide range of practical scenarios within a vehicular ISAC setting.}}} \cite{mishra2019toward}.

We now stack the observations in \eqref{eq_rec_bb2} over fast-time $\ell$ and slow-time $m$ to obtain the fast-time/slow-time radar observation matrix \cite{MIMO_OFDM_ICI_JSTSP_2021}:
\begin{align} \label{eq_ym_all_multi}
    {\YYr^{i}} = \underbrace{\DD(\cfo)}_{{\substack{\rm{CFO} } }}  \FF_N\herm \Bigg( \underbrace{{\PP \odot} \XX}_{{\substack{\rm{TX~Signal} } }}  \odot \overbrace{\bigg[ \sum_{k=0}^{K-1}   {\alpha_k^{i}} \,    \bb(\taubar_k) \cc\trp(\nubar_k) { \odot \underbrace{\boldone \atx\trp(\theta_k) \FF}_{{\substack{\rm{TX~Beam}\\\rm{Sweeping} } }}  } \bigg]}^{{\substack{\rm{Radar~Channel} } }} \Bigg)   + {\ZZr^{i}} \in \complexset{N}{M} ~,
\end{align}
where we assume $T \nu_k \ll 1, \, \forall k$ \cite{Passive_OFDM_2010,OFDM_DFRC_TSP_2021}. In \eqref{eq_ym_all_multi}, {$\FF = [\ffb_0 \, \ldots \, \ffb_{M-1}] \in \complexset{\Nt}{M} $ is the ISAC TX beamforming matrix, $\boldone$ represents an all-ones vector of conformant size,} $\DD(\cfo) = \diag{1, e^{j 2 \pi \frac{T}{N} \cfo}, \ldots, e^{j 2 \pi \frac{T(N-1)}{N} \cfo} } \in \complexset{N}{N}$ denotes the CFO-induced phase rotation matrix that captures the fast-time phase shifts within each symbol as a function of a given CFO/Doppler $\cfo$, $\bb(\tau) \in \complexset{N}{1}$ with $[\bb(\tau)]_n = e^{-j 2 \pi n \deltaf \tau} $ is the frequency-domain steering vector, $\cc(\nu) \in \complexset{M}{1}$ with $[\cc(\nu)]_m = e^{j 2 \pi m \Tsym \nu} $ is the time-domain steering vector, $\FF_N \in \complexset{N}{N}$ denotes the unitary DFT matrix with $\left[ \FF_N \right]_{\ell,n} = \frac{1}{\sqrt{N}} e^{- j 2 \pi n \frac{\ell}{N}} $, $\XX \in \complexset{N}{M}$ with $\left[ \XX \right]_{n,m} \triangleq \xnm$, {$\PP \in \realset{N}{M}$ with $\left[ \PP \right]_{n,m} \triangleq \sqrt{\pnm}$, and} $\YYr \in \complexset{N}{M}$ with $\left[ \YYr \right]_{\ell,m} \triangleq y_{\ell,m} $. {Moreover,} ${\ZZr^{i}} \in \complexset{N}{M}$ denotes {the disturbance term including clutter and} additive white Gaussian noise (AWGN) with $\vecc{{\ZZr^{i}}} \sim \mtCN(\boldzero, \allowbreak {\RR} ) $, {where $\RR \in \complexset{NM}{NM}$ is the covariance matrix of the disturbance\footnote{{For ease of exposition, we set $\RR = \sigmar^2 \Imatrix$ in the sequel. However, the presented methods and KPIs can be straightforwardly adapted to accommodate any arbitrary $\RR$.}}}. {In radar systems, clutter models a variety unwanted returns in addition to noise and interference in wireless communications. Clutter models many other propagation phenomena including atmospheric and weather effects, as well as radar countermeasures in military applications. It depends on the operational environment (sea, ground, air) with surface, volume and point clutter models, radar platform (ground-based, airborne,  grazing angle) and it can be also signal dependent, see \cite{GRECO2014513} for tutorial overview. Hence, detailed consideration of clutter is not feasible here.} Several remarks are in order regarding \eqref{eq_ym_all_multi}.
\begin{itemize}
    \item In \textbf{monostatic sensing}, $\DD(\cfo) = \Imatrix$, $\taubar_k = \tau_k$ and $\nubar_k = \nu_k$. Being co-located with the ISAC transmitter, the radar RX possesses complete knowledge of the transmit symbols $\XX$, including both data and pilots. Hence, the radar RX does not need to perform CFO compensation and can estimate the true delays and Doppler shifts of the existing targets without any clock bias and CFO, respectively, using the entire OFDM frame ${\YYr^{i}}$ (e.g., via 2-D FFT on $\FF_N {\YYr^{i}} \odot \PP \odot \XX^{*}$ {assuming constant TX beamforming across the symbols (no TX beam sweeping), i.e., $\ffb_0 = \ldots = \ffb_{M-1}$} \cite{Sturm11,Fan_ISAC_6G_JSAC_2022}). {In addition, the AoAs/AoDs $\theta_k$ can be estimated by processing $\{\YYr^{i}\}_{i=0}^{\Nss-1}$ along the spatial domain by exploiting the AoA/AoD-dependent phase variations in $\alpha_k^{i} = \alpha_k [\arx(\phi_k)]_i$ over $i=0,\ldots,\Nss-1$.}
    \item In \textbf{bistatic sensing}, the radar RX needs to estimate and compensate for the CFO, and can only estimate the delays $\tau_k$ and Doppler shifts $\nu_k$ up to a clock bias $\bias$ and CFO $\cfo$, using a fraction of the OFDM frame ${\YYr^{i}}$ due to partial knowledge of $\XX$ involving only pilots. Employing iterative data-aided channel estimation and data detection techniques could improve sensing performance by enabling the use of data decisions alongside known pilots to refine sensing channel estimates \cite{dataAided_bistatic_JCS_2022}. {Moreover, the AoAs $\phi_k$ can be estimated in the same way as in monostatic sensing, while the AoDs $\theta_k$ can be inferred by processing $\YYr^{i}$ along the time domain based on AoD-dependent phase changes in the beamspace observations $\boldone \atx\trp(\theta_k) \FF$.} 
\end{itemize}

\subsubsection{Communication Receive Signal Model}
Following the same steps leading to \eqref{eq_ym_all_multi}, the signal at the communications RX after time synchronization and CP removal (arranged into the fast-time/slow-time structure) can be written as \cite{MIMO_JCAS_OFDM_TWC_2023,OFDM_DFRC_TSP_2021}
\begin{align} \label{eq_y_com}
    \YYc = \DD(\cfoc) \FF_N\herm \big( {\PP \odot} \XX \odot \HHc \big)   + \ZZc \in \complexset{N}{M} ~,
\end{align}
where $\HHc = \sum_{k=0}^{K'-1} \alpha'_k \,    \bb(\tau'_k) \cc\trp(\nu'_k) { \odot \boldone \atx\trp(\theta'_k) \FF \odot \boldone \arxc\trp(\phi'_k) \VV } \in \complexset{N}{M}$ denotes the communication channel in the frequency-time domain with $K'$ paths, each characterized by a complex channel gain $\alpha'_k$, delay $\tau'_k$, Doppler shift $\nu'_k$, {AoD $\theta'_k$ and AoA $\phi'_k$, $\VV \in \complexset{\Nc}{M}$ is the communication RX beamforming matrix, $\arxc(\phi) \in \complexset{\Nc}{1}$ is the array steering vector at the communication RX, and} $\cfoc$ is the CFO between the ISAC TX and the communications RX. 
Additionally, $\ZZc \in \complexset{N}{M}$ is AWGN with $\vecc{\ZZc} \sim \mtCN(\boldzero, \allowbreak \sigmac^2 \Imatrix ) $. {Considering a simplified} scenario with {$\Nt = \Nc = 1$ and} small Doppler shifts (e.g., LOS-dominant mmWave vehicular channels with negligible Doppler shift between the transmitting and receiving vehicles moving in the same direction \cite{80211_Radar_TVT_2018}), 
$\HHc$ can be modeled as a quasi-static block-fading channel (i.e., frequency-selective, but time-invariant over a single block of $M$ symbols) \cite{OFDM_DFRC_TSP_2021}. In this case, the channel degenerates to $\HHc = \hhfreqc\boldone\trp$, where $ \hhfreqc = \sum_{k=0}^{K'-1} \alpha'_k \,    \bb(\tau'_k) $.

%%%%%%%%%%%%%%%%%%%%%%%%%%%%%%%%%%%%%%%%%%%%%%%%%%%%%%%
\subsection{Key Performance Indicators for OFDM Radar and Communications}
{This part focuses on radar and communication KPIs for OFDM ISAC systems. In general, these KPIs can be optimized by designing the power allocation matrix\footnote{{Subcarrier selection can be implemented as a special case in the design of $\PP$ 
by allocating a power level of $1$ to the selected subcarriers and $0$ to those that are not selected.}} $\PP$ \cite{MIMO_JCAS_OFDM_TWC_2023,OFDM_DFRC_TSP_2021}, as observed from \eqref{eq_ym_all_multi} and \eqref{eq_y_com}.}

\subsubsection{Radar KPIs}
Given the received signal $\YYr$ in \eqref{eq_ym_all_multi} and the transmit symbols $\XX$, the goal at the radar RX is to detect the presence of the $K$ targets and estimate their delay, Doppler {and angle} parameters. 
\paragraph{Detection}
The detection problem {at the $\thn{i}$ RX channel without TX beam sweeping} in the presence of a single target echo in \eqref{eq_ym_all_multi} (after CFO compensation in bistatic sensing) can be formulated as a binary composite hypothesis testing problem 
\begin{align}\label{eq_hypotest}
    {\YYr^{i}} = \begin{cases}
	{\ZZr^{i}},&~~ {\rm{under~\Hzero}}  \\
	\FF_N\herm \bigg( {\PP \odot} \XX \odot \alpha \, {[\arx(\phi)]_i} \,    \bb(\tau) \cc\trp(\nu) \bigg)   + {\ZZr^{i}},&~~ {\rm{under~\Hone}}  
	\end{cases} ~,
\end{align}
with the unknown target parameters $\alpha$, $\tau$, $\nu$ {and $\phi$} to be estimated, {where the hypotheses $\Hzero$ and $\Hone$ correspond to the absence and presence of a target.} 
To tackle \eqref{eq_hypotest}, a generalized likelihood ratio test (GLRT) \cite[Ch.~6.2.4]{richards2005fundamentals} can be derived, yielding as by-products the maximum-likelihood (ML) estimates of the unknown parameters\footnote{To cover the case of multiple targets, iterative interference cancellation procedures can be applied, where the strongest echo is detected at each iteration and its effect subtracted from $\YYr$ \cite{grossi2020adaptive,MIMO_OFDM_ICI_JSTSP_2021}.}. The probability of detection in \eqref{eq_hypotest} {considering all RX channels} can be approximated as \cite[Eq.~(6.52)]{richards2005fundamentals} $ \pd = Q_1\left( \sqrt{2 \gammasnr}, \sqrt{-2 \log \pfa} \right)$, 
 where $\gammasnr \triangleq \lvert \alpha \rvert^2 \normfsmall{{\PP \odot} \XX}  \Nss  / \sigmar^2$, $\pfa$ denotes the false alarm rate constraint for the GLRT detector and $Q_1(\cdot, \cdot)$ is the first-order Marcum Q-function. 

\paragraph{Estimation}
Target parameter estimation performance can be quantified by the Cram\'{e}r-Rao bound (CRB), {which is a function of $\PP$ \cite[Eq.~(S22)-(S23)]{OFDM_DFRC_TSP_2021}}. In the case of a single target, unit-amplitude symbols {and uniform power allocation} (i.e., $\lvert x_{m,n} \rvert = 1 \, {, \pnm = \ptot/(NM)} \,\, \forall m, n$) , the CRBs on range, velocity {and AoA/AoD} estimation in monostatic sensing {for a uniform linear array with element spacing $d$} are given by \cite{OFDM_DFRC_TSP_2021}
\begin{align} \label{eq_crbs}
    \Eee\{ (\widehat{R} - R)^2 \} \geq \frac{3 c^2}{8 \gammasnr  \pi^2 B^2} ~ , ~   \Eee\{ (\widehat{v} - v)^2 \} \geq \frac{3 c^2}{8 \gammasnr \pi^2 \fc^2 \Ttot^2} ~, { ~   \Eee\{ (\widehat{\phi} - \phi)^2 \} \geq \frac{3 \lambda^2}{2 \gammasnr \pi^2 D^2 \cos^2(\phi) } ~,}
\end{align}
where $B = {\deltaf \sqrt{N^2-1}} $, $\Ttot = {\Tsym \sqrt{M^2-1}} $ and {$D = d \sqrt{\Nss^2-1}$ represent, respectively, the approximate aperture sizes in frequency, time and space (i.e., the total bandwidth, the total duration and the array size)}, and $R = c\tau/2$ and $v = \lambda \nu/2$ denote the target range and velocity, respectively. When comparing detection performance and accuracy, it is important to note that the detection performance relies solely on the \ac{SNR} (excluding its inherent dependence on $\pfa$), while the estimation accuracy is affected by the \ac{SNR}, {the subcarrier power allocation $\PP$}, the bandwidth/duration of the OFDM frame for range/velocity estimation {and the array aperture (size) for angle estimation}.

\paragraph{Resolution}
In the presence of multiple targets, the range, velocity {and angle} resolution are given, respectively, by $c/(2B)$, $\lambda/(2 \Ttot)$ {and $0.89 \lambda/D$} \cite[Eq.~(1.9)]{richards2005fundamentals}. Combining this with \eqref{eq_crbs} suggests that large bandwidths, long frames {and large arrays} facilitate high-resolution and high-accuracy sensing {in MIMO-OFDM ISAC systems}.

\paragraph{{Other KPIs}}
{Additional radar KPIs for MIMO-OFDM ISAC systems can include mutual information (MI) between the radar channel and the received signal in \eqref{eq_ym_all_multi}, along with the beampattern matching error relative to a desired radar beampattern \cite{overview_JCRS_2021}. These KPIs can be optimized through the design of $\PP$ and/or $\FF$.}

\subsubsection{Communication KPI}
Given $\YYc$ in \eqref{eq_y_com} and assuming the channel $\HHc$ and the CFO $\cfoc$ are estimated a-priori via pilot symbols, the goal of the communications RX is to decode the transmit symbols $\XX$. The achievable rate is often employed as the communication KPI:
\begin{align} \label{eq_rate}
    C = \sum_{n=0}^{N-1} \sum_{m=0}^{M-1} \log\left( 1 + \frac{ {\pnm} \lvert[\HHc]_{n,m}\rvert^2 }{\sigmac^2} \right) ~.
\end{align}
Optimizing ISAC performance in OFDM systems {via the design of $\PP$} involves inherent trade-offs between communication and radar KPIs, particularly in terms of estimation accuracy. A notable example is water-filling power allocation {for $\PP$}, which maximizes the rate in \eqref{eq_rate}. However, this strategy may lead to suboptimal accuracy, i.e., high CRB, since minimizing CRB implies maximizing the root mean squared (RMS) bandwidth in delay estimation, which necessitates spreading the power towards the edges of the spectrum \cite{OFDM_DFRC_TSP_2021}.

%%%%%%%%%%%%%%%%%%%%%%%%%%%%%%%%%%%%%%%%%%%%%%%%%%%%%%%
\subsection{Multicarrier Phase-Coded Radar Waveforms}

The MCPC radar waveforms \cite{levanon2000multifrequency} modulate each subcarrier by a code sequence while maintaining the orthogonality of subcarriers. This corresponds to time-domain spreading on each subcarrier after serial to parallel conversion, hence it is closely related to MC DS-CDMA (multicarrier direct-sequence CDMA) used in communications systems. 
The codes may form a complementary set, 
and the code design may also consider the ambiguity function (AF) in delay and Doppler domains, and the peak-to-average power ratio (PAPR) for  efficient use of amplifiers.  If multiple waveforms are launched simultaneously,
cross-correlation properties need to be taken into account. In addition, waveform designs and adaptation are typically different for different radar sensing tasks and target scenarios.  

The phase coding for MC radar modulates each subcarrier by a code sequence of a specific length $L$ such that 
the subcarriers remain orthogonal. Hence, the signal will occupy a broader bandwidth. To maintain orthogonality, the intercarrier spacing for an MCPC waveform needs to accommodate the increased bandwidth. Examples of spreading codes include P3 or P4 polyphase codes, Barker, Zadoff-Chu, Kasami and Gold codes. 
The selection of codes and their allocation on subcarriers impact the ambiguity function (AF) of the waveform. 
The use of CP is necessary for the MC communications tasks to make the receiver processing simple whereas for radar sensing it is not necessary.  The CP may induce unwanted correlations at certain delays, thus masking targets at certain distances. 
However, a monostatic radar may require guard periods and repeating pulses at some frequency to avoid range ambiguities.

\begin{figure}[t!] 
	\centering
	\includegraphics[scale=0.7]{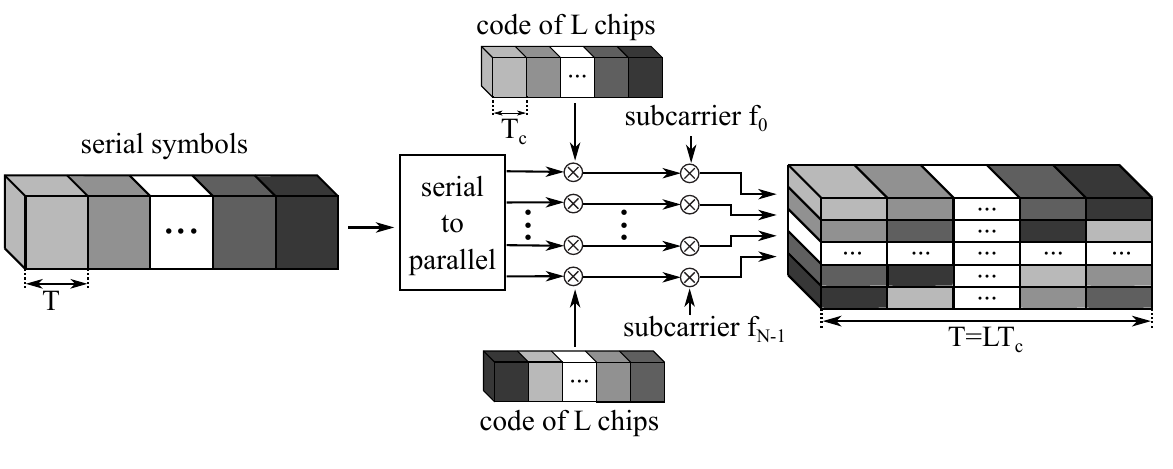}
	\vspace{-4mm}
     \caption{Generation of a $N \times L$ MCPC pulse comprised of $N$ subcarriers modulated by phase codes of length $L$. The intercarrier spacing is ${1}/{T_c}$ and the overall symbol duration is $T = LT_c$. Intercarrier spacing is increased to maintain orthogonality when spreading takes place.}
     \vspace{-3mm}
	\label{fig:mcpc_structure}
\end{figure}

The complex envelope of an $N\times L$ MCPC waveform is given by:
\begin{equation} 
\label{MCPC_complex_envelope}
g(t)=\sum^{N-1}_{n=0}\sum^{L-1}_{l=0} w_n c_{n,l} s[t-(l-1)T_c] \exp \left [j 2\pi \left(n-\frac{N+1}{2} \right)\frac{t}{T_c} \right],  \quad 0\leq t \leq LT_c
\end{equation}
where $c_{n,l}$ is the $l$th element of the phase code sequence modulating the $n$th subcarrier, $w_n$ is the complex weight associated with the $n$th subcarrier and $s(t) = \rect{t} = 1$ for $0 \leq t \leq T_c$ and zero elsewhere. 
The complex weight $w_n$ can be the same for all subcarriers or used to adjust power on each subcarrier, for example to reduce the PAPR or adapt to channel quality or particular sensing task requirements. Such weighting, however, would impact the AF of the waveform. The overall duration of a MCPC pulse is $T=LT_c$, with the duration of one chip of $T_c$. The intercarrier spacing is increased to $1/T_c$ and the effective bandwidth is $N/T_c$. The spectrum of an MCPC pulse is relatively flat. 
The autocorrelation main lobe width of such signal is $T_c/N$, while the AF main lobe width in Doppler domain is $1/T$. 
The generation of MCPC waveform is depicted in Fig. \ref{fig:mcpc_structure}.

The main difference between MCPC and OFDM waveforms with equal
bandwidth and duration is that 
MCPC has fewer subcarriers with larger spacing. When $L=1$, we have a conventional OFDM waveform. MCPC and spreading provide a higher range resolution, resistance to multipath, frequency diversity, low probability of intercept, {low probability of detection} (LPI/LPD) properties and improved immunity to narrowband interferences compared to OFDM. A generalized multicarrier radar (GMR) model that describes most well-known waveforms using a compact matrix notation was introduced in \cite{bicatrsp16}. Frequency hopping, step approximation of {linear frequency modulation} (LFM), OFDM and MCPC waveforms are special cases of the GMR model. 
Commonly, in multicarrier systems, resource blocks of multiple subcarriers are often used, for example 12 consecutive subcarriers. They may be designated to different sensing and communications tasks. 

%%%%%%%%%%%%%%%%%%%%%%%%%%%%%%%%%%%%%%%%%%%%%%%%%%%%%%%
\subsection{{Emerging Multicarrier Waveforms Towards 6G}}

OFDM, discrete Fourier transform spread OFDM (DFT-s-OFDM),  orthogonal time frequency space (OTFS), and orthogonal time sequency multiplexing (OTSM)  are considered potential waveform candidates for 6G wireless communication and sensing. 
Each of these waveforms has unique benefits and drawbacks for communication and sensing applications. 
OFDM has been extensively used in earlier generation networks due to its high spectral efficiency. It has the ability to handle severe channel conditions without complex equalization filters. OFDM's architecture allows it to be easily adaptable to MIMO systems, which is crucial for achieving high data rates. On the other hand, it is 
sensitive to two impairments: Doppler shift, which could be a challenge in high mobility scenarios, and power amplifier nonlinearity, due to the high PAPR. 
These limitations are, respectively, addressed by DFT-s-OFDM and OTFS. 
DFT-s-OFDM involves precoding an OFDM signal to improve the power efficiency and sensitivity to power amplifier nonlinearity, especially in uplink scenarios where coverage is limited. 
DFT-s-OFDM is supported in the uplink of 4G and 5G, and the same is very likely to apply also in 6G. The potential role of DFT-s-OFDM in 6G downlink, as an energy-saving and/or improved coverage feature, is currently open. 
OTFS is gaining traction as a promising waveform for 6G due to its two-dimensional modulation format, which is highly resilient to channel impairments and provides enhanced performance in high mobility scenarios \cite{otfs_vs_ofdm_2022}. OTFS can also be interpreted as a pre-coded version of OFDM, but without a cyclic prefix between each symbol. The limitations of OTFS relate to its high processing complexity, which is not easy to overcome as they are inherent to its on-grid 2D convolution method. 
The recently introduced single-carrier OTSM scheme offers a lower-complexity alternative to OTFS while attaining comparable rate performance \cite{otsm}.

\section{MC-ISAC under Hardware Non-idealities}

Signal processing in MC-ISAC demands higher accuracy than communication-only systems, increasing complexity and overhead. Radar sensing estimates a structured geometric channel, typically a nonlinear function of physical parameters, unlike communication channel estimation which often involves linear problems with an unstructured channel. {Illuminated radar targets and their radar cross-sections (RCS) are important parts of the propagation phenomena and sensing channels. In radar sensing, clutter is caused by a variety of propagation effects. It is observed as different kinds of interference at the receiver. Clutter modeling and mitigation are thoroughly investigated topics in radar research \cite{GRECO2014513}.} Moreover, real-world imperfections in analog/RF hardware can degrade the channel estimation performance. This section discusses hardware/channel non-idealities in MC-ISAC systems, focusing on their impact as well as strategies for mitigation and exploitation.

\subsection{Impact of Non-Idealities} 
The most notable non-idealities are the transmitter power amplifier (PA) nonlinearities and the phase noise in the RF oscillators. Moreover, inter-carrier interference poses a formidable challenge for OFDM. In antenna array systems, various coupling effects and other perturbations may take place.

\subsubsection{Power Amplifier Nonlinearity}
For modeling of the PA nonlinear distortion, there exist generally a plethora of alternative models ranging from instantaneous polynomials and look-up tables to more advanced models such as the Volterra series and its different subsets such as the widely adopted memory polynomial (MP) and generalized memory polynomial (GMP){\cite{GhanMM09}}. A common feature in all the models, and in the physical PA behavior, is that the level of distortion depends on the input signal envelope behavior and particularly on the PAPR. For example, the MP model can be expressed as $x_{\text{MP}}(n) = \sum_{\substack{p=1 \\ p \text{ odd}}}^{P}\sum_{\substack{m=0}}^{M} a_{p,m} x(n-m)|x(n-m)|^{p-1}$, 
 with order  $P$ and memory $M$, where $x(n)$ denotes the complex baseband equivalent of the PA input while $|x(n)|$ is the corresponding envelope.

\subsubsection{Inter-Carrier Interference}
With high-speed targets and/or small $\deltaf$, the radar signal model \eqref{eq_ym_all_multi} is no longer valid since the assumption $T \nu_k \ll 1, \, \forall k$ (stated after \eqref{eq_ym_all_multi}) is violated. A more general model including the effect of inter-carrier interference (ICI) for each target due to high-mobility and/or small $\deltaf$ can be written as{\footnote{{For simplicity of illustration, we consider a single-antenna TX and a single-antenna radar RX in this part and the next part on phase noise.}}} \cite{MIMO_OFDM_ICI_JSTSP_2021}
\begin{align} \label{eq_y_ici}
    \YYr = \sum_{k=0}^{K-1} \alpha_k  \underbrace{\DD(\nubar_k)}_{\substack{\rm{ICI} } } \FF_N\herm \Big({\PP \odot} \XX \odot \bb(\taubar_k) \cc\trp(\nubar_k) \Big)  + \ZZr ~.
\end{align}
Focusing on monostatic sensing for ease of exposition, the maximum phase excursion (MPE) in $\DD(\nu)$ is given by $2 \pi T \nu =  4 \pi v \fc / (c \deltaf) $, where $v = \lambda \nu /2$ is the target speed in $\rm{m/s}$. For standard 5G NR FR2 parameters with $\deltaf = 120 \, \rm{kHz}$ and $\fc = 28 \, \rm{GHz}$, 
a target with $v = 20 \, \rm{m/s}$ leads to an MPE of $0.196$, which is much smaller than $2 \pi$, and thus $\DD(\nu)$ can be approximated as an identity matrix, in which case \eqref{eq_y_ici} degenerates to \eqref{eq_ym_all_multi}. When $\deltaf = 60 \, \rm{kHz}$ and $v = 80 \, \rm{m/s}$ (e.g., two cars approaching each other with $144 \, \rm{km/h}$), the MPE becomes $1.563$, which is comparable to $2 \pi$ and cannot be neglected. Fig.~\ref{fig_ici_range_profile} demonstrates the impact of ICI in high-mobility scenarios on the range profile obtained via standard 2-D FFT processing \cite{Sturm11,Fan_ISAC_6G_JSAC_2022} on $\YYr$ in \eqref{eq_y_ici}. {The impact of ICI becomes more severe as $\fc$ increases since Doppler shift is proportional to $\fc$.}

\subsubsection{Phase Noise}
For an arbitrary baseband equivalent signal $x(n)$, the fundamental behavioral model of oscillator phase noise (PN) reads ${x_\text{PN}}(n)=x(n)e^{j\phi_\text{PN}(n)}$ where $\phi_\text{PN}(n)$ refers to the PN. Hence, PN is seen as time-varying random phase fluctuations that may vary considerably within an individual multicarrier symbol duration. This causes spreading of the waveform spectrum whose out of band part can cause adjacent channel interference while the passband part corresponds to the ICI. PN happens in both transmitting and receiving entities, while the user equipment (UE) PN is typically larger than that of a gNB.

\begin{figure}
    \centering
    \begin{minipage}{0.48\textwidth}
        \centering
        \includegraphics[width=\linewidth]{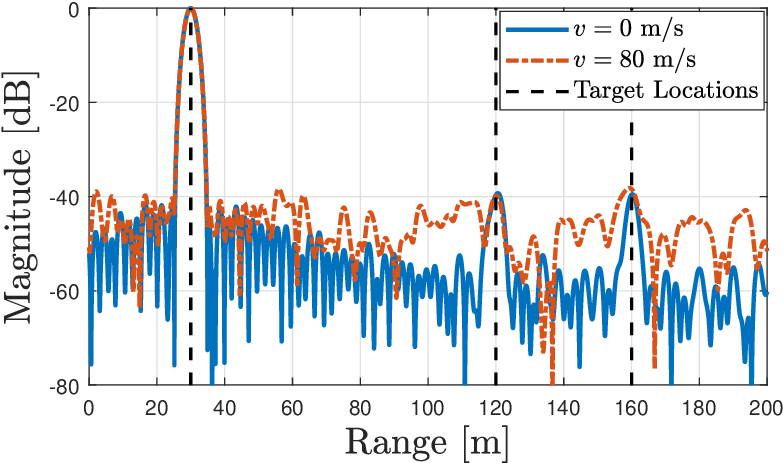}
        \caption{Impact of ICI on the range profile of OFDM radar with the parameters $\fc = 28 \, \rm{GHz}$, $\deltaf = 60 \, \rm{kHz}$, $N = 1024$ and $M = 16$, in the presence of $3$ targets having the same velocity $v$, the ranges $(30, 120, 160) \, \rm{m}$ and the SNRs (quantified by $\lvert\alpha_k\rvert^2/\sigmar^2$ in \eqref{eq_y_ici}) $(30, -10, -10) \, \rm{dB}$, respectively. {The range profiles are obtained via standard 2-D FFT processing \cite{Sturm11,Fan_ISAC_6G_JSAC_2022} on \eqref{eq_y_ici}.} In scenarios characterized by high mobility, ICI results in elevated side-lobe levels and can obscure weaker targets, leading to missed detections.}
        \label{fig_ici_range_profile}
    \end{minipage}
    \hspace{0.2cm}
    \begin{minipage}{0.48\textwidth}
        \centering
        \includegraphics[width=\linewidth]{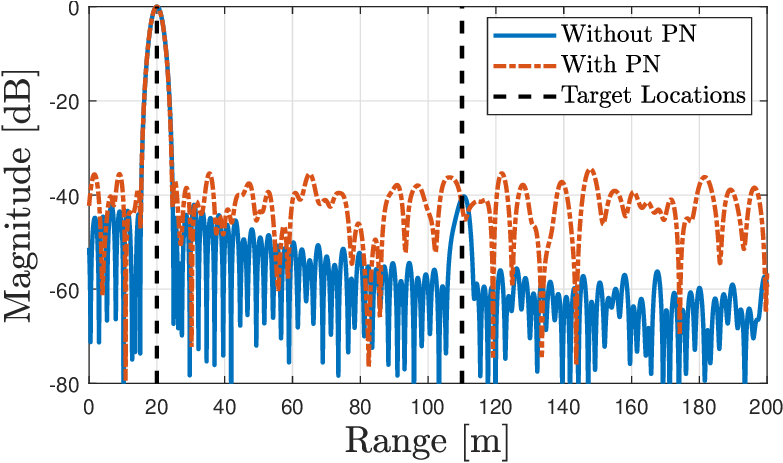}
        \caption{Impact of PN on the range profile of OFDM radar with the parameters $\fc = 28 \, \rm{GHz}$, $\deltaf = 120 \, \rm{kHz}$, $N = 512$ and $M = 8$, in the presence of $2$ targets with the ranges $(20, 110) \, \rm{m}$, the velocities $(0, 0) \, \rm{m/s}$ and the SNRs (quantified by $\lvert\alpha_k\rvert^2/\sigmar^2$ in \eqref{eq_y_pn}) $(30, -5) \, \rm{dB}$, respectively. The oscillator is a free-running oscillator with the $3 \,\rm{dB}$ bandwidth $100 \, \rm{kHz}$. {The range profiles are obtained via standard 2-D FFT processing \cite{Sturm11,Fan_ISAC_6G_JSAC_2022} on \eqref{eq_y_pn}.} PN leads to increased side-lobe levels and can mask weaker targets.}
        \label{fig_pn_range_profile}
    \end{minipage}
    \vspace{-0.2in}
\end{figure}

In the presence of oscillator PN, the radar signal model \eqref{eq_ym_all_multi} should be generalized to involve a multiplicative PN component in the time domain. For the special case of monostatic sensing, \eqref{eq_ym_all_multi} becomes {\cite{PN_Exploitation_TSP_2023}}
\begin{align} \label{eq_y_pn}
    \YYr = \sum_{k=0}^{K-1} \alpha_k  \underbrace{\WW(\tau_k)}_{\substack{\rm{PN} } } \odot\, \FF_N\herm \Big({\PP \odot} \XX \odot \bb(\tau_k) \cc\trp(\nu_k) \Big)  + \ZZr ~,
\end{align}
where $\WW(\tau_k) \in \complexset{N}{M}$ denotes the multiplicative PN matrix associated with the $\thn{k}$ target and contains fast-time/slow-time samples from the self-referenced PN process at the monostatic radar RX. With a slight abuse of notation, $\WW(\tau_k)$ does not represent a deterministic function of $\tau_k$; rather, it is used to indicate that the statistics of $\WW(\tau_k)$ depend on $\tau_k$. In Fig.~\ref{fig_pn_range_profile}, we showcase the impact of PN by plotting the range profiles (obtained via 2-D FFT on $\YYr$ in \eqref{eq_y_pn}) for both ideal and non-ideal oscillators. {The detrimental effects of PN are expected to be less pronounced in 5G FR1 compared to 5G FR2 as the severity of PN decreases with decreasing $\fc$ \cite{PN_Exploitation_TSP_2023}.}

\subsubsection{Self-Interference (SI)}\label{sec_si}
We consider now the effect of SI that appears in MIMO-OFDM when operating in a monostatic setting {\cite{BF_FD_JCAS_TCOM_2022,Bayraktar2024}}. At the transmitter side of the MIMO-OFDM ISAC transceiver, the number of antenna elements is $\Ntx$, while at the receiver side the system operates with $\Nrx$ antennas. The response of the antenna arrays at a given angular direction is described by the array steering vectors, denoted as $\atx(\theta) \in \complexset{\Ntx}{1}$ for the transmitter and $\arx(\phi) \in \complexset{\Nrx}{1}$ for the receiver, with $\theta$ the angle of arrival and $\phi$ the angle of departure.  The transmitted signal 
includes $\Ns$ streams of OFDM symbols. We assume that the number of streams is equal to the number of transmit RF chains  available at the transmitter, denoted as $\Lt$. For simplicity, we also assume that the number of RF chains at the receiver, $\Lr$, is equal to the number of streams. If the system operates at sub-6 GHz, it is possible to assume that a fully digital MIMO architecture can be used, i.e. there is one RF chain per antenna. In this case, the precoder and the combiner used in the MIMO-OFDM transceiver at subcarrier  $n$ and symbol $m$, denoted as $\bF[n,m] \in \complexset{\Ntx}{\Ns} $ and $\bW[n,m] \in   \complexset{\Nrx}{\Ns} $, are fully digital matrices. When operating at mmWave frequencies, the precoders and combiners are implemented with a hybrid MIMO architecture which splits the spatial processing into an analog and a digital stage. In other words, $\bF=\Fbb[n,m]\Frf[{m}]$, and $\bW=\Wbb[n,m]\Wrf[{m}]$, with  $\Fbb[n,m] \in \complexset{\Lt}{\Ns}$  the digital baseband precoder at subcarrier $n$ and symbol $m$, $\Frf[{m}] \in \complexset{\Ntx}{\Lt}$  the frequency flat analog precoding matrix applied at symbol $m$,  $\Wbb[n,m] \in \complexset{\Ns}{\Lr}$ the digital baseband combiner,  and  $\Wrf[{m}] \in \complexset{\Lr}{\Nrx}$ the analog combiner. {Note that the specific implementation of the analog precoding/combining matrices will introduce additional constraints in the feasible values for $\Frf[{m}]$ / $\Wrf[{m}]$. For example, if a phase shifting network is used to implement the analog stage, a unit modulus constraint should be considered. Alternative implementations based on attenuators or variable gain amplifiers (VGA) may impose other constraints on the feasible values for the magnitudes.}

If we denote by $\tilde{\mathbf{x}}_{m,n}$ the vector containing the OFDM symbol for all the streams at symbol index $m$ and subcarrier $n$, and assuming that the number of point targets in the environment is $K$, the corresponding received signal at the output of the analog combiner in the ISAC transceiver can be written as {\cite{BF_FD_JCAS_TCOM_2022,Bayraktar2024}}
 \begin{equation}\label{eq:MIMO-OFDM}
    \by_{n,m} = \Wrf[{m}]\herm \Hrnm \Frf[{m}] \Fbb[n,m] \tildex_{n,m} + \underbrace{\Wrf[{m}]\herm \bH_{\rm{SI}} \Frf[{m}] \Fbb[n,m] \tildex_{n,m} }_{\text{SI}} + \bz_{n,m} \in \complexset{\Lr}{1} {\,,}
\end{equation}
where $\Hrnm \in \complexset{\Nrx}{\Ntx}$ represents the radar channel at symbol index $m$ and subcarrier $n$, while $\bH_{\rm{SI}} \in \complexset{\Nrx}{Ntx}$ is the SI channel, usually modeled as a frequency flat line-of-sight (LOS) channel which represents the coupling between the transmit and receive arrays of the monostatic ISAC transceiver. Note that it remains the same for different OFDM symbols. 
The radar channel is given by $\Hrnm=\AR(\boldsymbol \theta)\Delta_{n,m}\AT\herm(\boldsymbol \theta)$, 
where $\Delta_{n,m}  \in \complexset{K}{K}$ is a diagonal matrix with $\Delta_{n,m}\vert_{k,k}=\alpha_k e^{-j 2 \pi n \deltaf \tau_k}  e^{j 2 \pi m \Tsym \nu_k  }$, while 
$\AR(\boldsymbol \theta)=[\arx(\theta_1), . . . , \arx(\theta_K)]$ and $\AT(\boldsymbol \theta)=[\atx(\theta_1), . . . , \atx(\theta_K)]$ contain the steering vectors at the receiver and transmitter side, respectively,  of the ISAC transceiver evaluated at the directions of the $K$ targets. It is assumed that the directions of departure and arrival are the same, since the transmit and received arrays are colocated. 

\subsubsection{Antenna Array Impairments}
When operating in a MIMO or phased array setting some impairments associated to the antenna array may appear, distorting the expected radiation pattern. 
First, we can consider {\em mutual coupling} between antenna elements, both at the transmit and the receive arrays, which can be represented by the matrices $\mathbf{C}_\text{T}$ and 
$\mathbf{C}_\text{R}$. Second, for a more practical representation of the antenna array behavior we can also consider the {\em calibration error}, represented by the diagonal matrices $\boldsymbol{\Gamma}_\text{T}$ and $\boldsymbol{\Gamma}_\text{R}$, which contain 
antenna gain and phase errors for the transmit and received arrays. Finally, {\em errors in the phase centers and spacings among antenna elements} due to manufacture errors also have to be considered. This impacts the expression of the array steering vectors, which no longer are computed assuming a uniform separation between elements. As a result of these impairments, the expression of the MIMO channel matrix accounting for impairments becomes {\cite{Xie2020TWC}}
\begin{equation}
\tilde{\mathbf{H}}_{n,m}=\mathbf{C}_\text{R}\boldsymbol{\Gamma}_\text{R}\tilde{\mathbf{A}}_\text{R}\boldsymbol{\Delta}_{n,m}\tilde{\mathbf{A}}_\text{T}^{*}\boldsymbol{\Gamma}_\text{T}^*\mathbf{C}_\text{T}^*,
\end{equation}
where $\tilde{\mathbf{A}}_\text{R}$ and $\tilde{\mathbf{A}}_\text{R}$ are the matrices that contain the perturbed steering vectors for the transmitter and receiver respectively, evaluated at the actual angles of arrival (AoA) and angles of departure (AoD), while
the diagonal matrix $\boldsymbol{\Delta}_{n,m}$ contains the complex gains for the different channel paths.
 
%%%%%%%%%%%%%%%%%%%%%%%%%%%%%

\subsection{Mitigation of Non-Idealities}

\subsubsection{Mitigating ICI and PN}
To mitigate the impact of intercarrier interference (ICI) in \eqref{eq_y_ici}, the task of delay-Doppler estimation can be framed as a joint CFO-channel estimation problem and an orthogonal matching pursuit (OMP)-like interference cancellation procedure can be employed in an iterative fashion to detect multiple targets \cite{MIMO_OFDM_ICI_JSTSP_2021}. Meanwhile, for mitigating the effects of PN in \eqref{eq_y_pn}, a hybrid ML/MAP estimator, coupled with an iterated small angle approximation approach, offers a near-optimal solution for PN compensation and range estimation \cite{PN_Exploitation_TSP_2023}. {Mitigation of ICI and PN for sensing in a monostatic ISAC system differs from a communications-only setup in two critical aspects: \textit{(i)} the entire OFDM frame is utilized as pilots for sensing, and \textit{(ii)} the approach involves estimating individual target parameters within a structured geometric radar channel, rather than an unstructured composite communication channel. An additional aspect specific to PN mitigation concerns the delay-dependency of the PN statistics in monostatic sensing as opposed to delay-independent PN statistics in communications.}

\subsubsection{Spatial Design For Self-Interference Mitigation}
The design of the precoders and combiners in \eqref{eq:MIMO-OFDM} has to tackle three different objectives:  suppress the self-interference (SI) term in \eqref{eq:MIMO-OFDM} to avoid saturation of low noise amplifiers (LNA) and analog-to-digital converters (ADC), illuminate the target either for detection or tracking, and enable downlink communication. One possible way to consider communication and radar performance in the design of the spatial filters for the monostatic ISAC transceiver is to maximize the achievable rate per subcarrier, which in the MIMO-OFDM case also depends on the precoders $\Fbb[n,m]$ and $\Frf[n]$ in \eqref{eq:MIMO-OFDM} and the combiners at the users, while guaranteeing a minimum gain in the directions of the targets. 
 For a given target angle $\theta_{\rm r}$, the transmit gain can be written as
\begin{equation}\label{eq_transmit_target_gain}
      G_{\mathrm{T}, n_s, n}(\theta_{\rm r}) = \left|\atx\herm (\theta_{\rm r}) \left[\Frf \Fbb[n,m]\right]_{:,n_s} \right|^2,
      \end{equation}
      for $n_s = 1,\dots,\Ns$. 
The receive target gain can be defined in  a similar manner from the receive steering vector $\arx (\theta_{\rm r})$.
Moreover, to mitigate SI, the precoders and combiners should satisfy
\begin{equation}\label{eq_SI}
      \Wrf\herm \bH^{\rm{SI}}_{n} \Frf \Fbb[n,m] = \boldzero_{\Lr \times \Ns}.
 \end{equation}
 
Some designs in the recent literature have addressed the precoder/combiner design {in different ways \cite{LiyanaarachchiTWC2024,Bayraktar2024,Bayraktar2023self,BF_FD_JCAS_TCOM_2022}}. For example, \cite{Bayraktar2023self} proposes the design of 
a  beam codebook for the analog precoding and combining stages of the monostatic ISAC transceiver which minimizes the SI, while guaranteeing a minimum gain on a grid of potential angles for the users and targets and the unit magnitude of the beamforming weights, so they can be implemented with phase shifters. The approach in \cite{BF_FD_JCAS_TCOM_2022} achieves the SI cancellation 
through the design of $\Wrf$, which also provides a high gain in the direction of the target. On the transmit side,  a multibeam hybrid precoder  provides sufficient gain at the communication and radar directions at the same time. 
A more general system model may be considered where the \ac{DL} and radar channels can have multiple paths which do not necessarily overlap {\cite{Bayraktar2024}}. On one side, the proposed design for the precoder maximizes the achievable rate and guarantees a sufficient TX target gain. On the other side, both the precoder and combiner contribute to the  SI  mitigation, and  the combiner also keeps the RX target gain above a threshold.

\subsubsection{Dictionary Learning For Mitigating Antenna Array Non-Idealities}
To mitigate the impact of antenna array impairments it is possible to define a dictionary learning process which implicitly calibrates the array impairments at both ends \cite{Xie2020TWC}. 
This way,  the impairments are included into the sparsifying dictionary used to represent the {frequency selective} channel. Then, the channel estimation stage {exploits the OFDM waveform} to compute the channel parameters assuming the learned dictionaries with impairments, acting effectively as a joint calibration and parameter estimation procedure, without need of conventional calibration.

\subsection{Exploitation of Non-Idealities}
Departing from the traditional approach of treating ICI and PN effects solely as
impairments in OFDM ISAC systems, we show how these effects can be exploited to improve OFDM monostatic sensing performance. The key observation enabling this exploitation in monostatic sensing as opposed to bistatic sensing and communications is that the bistatic sensing/communications RX employs an \textit{independent} oscillator, while the monostatic sensing RX uses the \textit{shared} oscillator with the ISAC TX for downconversion. The implications of this are elaborated on in the following.

\subsubsection{Resolving Doppler Ambiguity via ICI Exploitation}
\begin{figure}
    \centering
    \begin{minipage}{0.45\textwidth}
        \centering
        \includegraphics[width=0.8\linewidth]{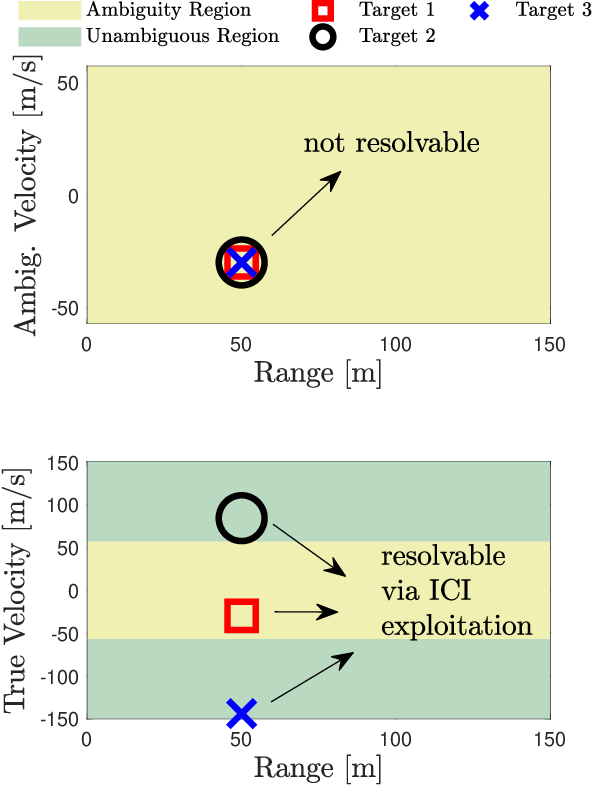}
        \caption{ICI exploitation to resolve Doppler ambiguity of high-speed targets and enhance target separability/resolvability via the use of additional Doppler information in $\DD(\nu)$ in \eqref{eq_y_ici}. The OFDM parameters are $\fc = 60 \, \rm{GHz}$, $N = 512$, $B = 100 \, \rm{MHz}$ and $M = 16$.}
        \label{fig_ici_exp}
    \end{minipage}
    \hspace{0.2cm}
    \begin{minipage}{0.45\textwidth}
        \centering
        \includegraphics[width=\linewidth]{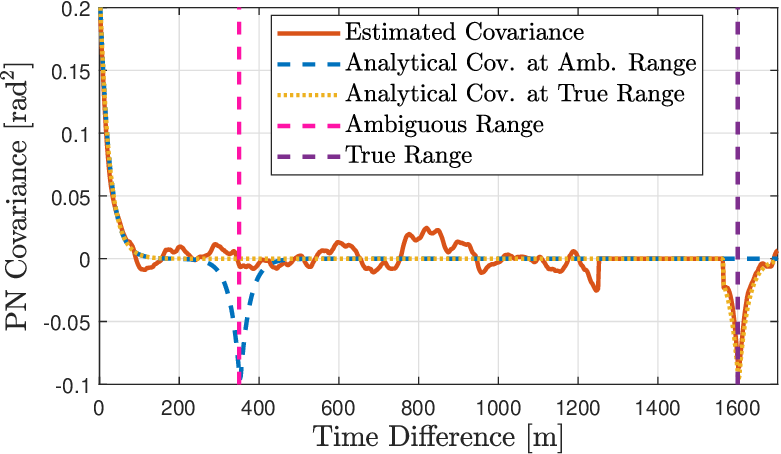}
        \caption{PN exploitation for resolving range ambiguity in monostatic sensing via covariance matching between the analytical Toeplitz-block-Toeplitz covariance matrices of the PN evaluated at candidate ranges and the empirical one obtained from the estimates of the PN matrix $\WW(\tau)$ in \eqref{eq_y_pn}. The OFDM parameters are $\fc = 28 \, \rm{GHz}$, $\deltaf = 120 \, \rm{kHz}$, $N = 512$ and $M = 10$, while the oscillator is a phase-locked loop (PLL) synthesizer with the loop bandwidth $1 \, \rm{MHz}$ and the $3 \,\rm{dB}$ bandwidth $20 \, \rm{kHz}$.}
        \label{fig_pn_exp}
    \end{minipage}
    \vspace{-0.22in}
\end{figure}

At the bistatic sensing/communications RX, both the Doppler shifts $\nu_k$ resulting from the mobility of TX, RX, and scatterers, and the CFO between the oscillators of TX and RX contribute to the ICI effect. This can be observed from $\DD(\nubar_k)$ in \eqref{eq_y_ici}, where $\nubar_k = \nu_k + \cfo$. In contrast, ICI at the monostatic sensing RX only arises from Doppler effects, as the same oscillator is utilized for both upconversion and downconversion. Consequently, the Doppler information contained in the ICI term $\DD(\nu)$ in \eqref{eq_y_ici} offers additional information into target velocity, with a significantly larger unambiguous detection limit compared to the slow-time Doppler information in $\cc(\nu)$ \cite{MIMO_OFDM_ICI_JSTSP_2021}. This is because the time-domain sampling frequency in $\DD(\nu)$ is $N$ times higher than that in $\cc(\nu)$. Fig. \ref{fig_ici_exp} illustrates how this property can be exploited to estimate the true velocities of high-mobility targets and differentiate among targets that would otherwise share the same range-Doppler bin. {This leads to higher accuracy in velocity estimation and improves multi-target detection performance.}

\subsubsection{Resolving Range Ambiguity via PN Exploitation}
Due to the use of the same oscillator for upconversion and downconversion in monostatic sensing, the statistics of the PN $\WW(\tau)$ in \eqref{eq_y_pn} carry delay information through its delay-dependent covariance matrix \cite{PN_Exploitation_TSP_2023}. Hence, this additional range information beyond what is provided by frequency-domain phase shifts in $\bb(\tau)$ can be leveraged to enhance ranging performance. Additionally, the range information in the statistics of $\WW(\tau)$ does not suffer from ambiguity limits as in $\bb(\tau)$, which enables resolving range ambiguity of far-away targets. Fig.~\ref{fig_pn_exp} illustrates how a covariance matching approach can allow us to estimate the true range of a far-away target. {Hence, PN exploitation can provide improved ranging accuracy for distant targets.}

\section{Multicarrier-ISAC Optimization {and Learning}}

Structured optimization and machine learning are key technologies in co-designing and jointly optimizing the performance of sensing and communication tasks in ISAC for mutual benefit.
{We will consider these two different approaches for efficient co-design and resource allocation in ISAC.
Structured optimization requires accurate information on the model, operational environment, parameters and constraints of the ISAC system.  
However, there may be a substantial modeling deficit that can be addressed via data-driven machine or statistical learning. Moreover, the nature of the problem and its constraints do not lend themselves to commonly used optimization tools or algorithms. Machine learning may provide algorithms to solve such problems.}

{The ISAC problem under consideration can be expressed as a constrained multi-objective optimization \cite{overview_JCRS_2021}
\begin{subequations}
\label{eq:pareto}
\begin{align}
    \min_{R} & \quad (f_{\text{comm}}(R),f_{\text{sense}}(R))\\
    \text{s.t.} & \quad  \mathbf{h}(R) \le \mathbf{0},
\end{align}
\end{subequations}
where $R$ represents the resources (e.g., power allocation $\mathbf{P}$, {subcarrier selection, beam or beamformer selection,} or sequence of precoders $\mathbf{F}$ in \eqref{eq_ym_all_multi}), $f_{\text{comm}}(R)$ represents a communication metric (e.g., achievable rate \eqref{eq_rate}), $f_{\text{sense}}(R)$ represents a sensing metric (e.g., the CRB \eqref{eq_crbs}), while $\mathbf{h}(R) \le \mathbf{0}$ \rev{represents the constraints, such as power constraints,  similarity constraints of waveforms or ambiguity function shape, {desired SINR levels}, or unit-modulus constraints for analog beamforming {or efficient use of amplifiers}. The constraints and their dimensions are often task and system specific, hence they are not specified explicitly for the general model above. To develop solutions to \eqref{eq:pareto}, scalarization by combining the objectives linearly and associating a  weight with each sub-objective, moving some objectives as constraints or finding a Pareto solution in which improvement in one objective cannot be done without worsening the other objectives
are widely used techniques \cite{overview_JCRS_2021}. }
}

\vspace{-0.1in}
\subsection{Structured ISAC Optimization}

{Structured optimization methods assume that all necessary model parameters needed in optimization are known or estimated reliably without uncertainty. In ISAC, they rely on ideal models and assumptions about the state of the radio environment, acquired data, hardware, and user and target scenarios.  Modeling a large-scale distributed radio system with dynamic propagation environments and spectrum usage patterns using realistic and rigorous mathematical modeling may not be feasible in practice. }Typically methods employ batch algorithms and may not be suitable for dynamic scenarios.  Moreover, no learning from the past experiences takes place.  

The optimization problem may have \emph{sensing-centric} or \emph{communication-centric} formulation. In sensing-centric designs,  performance criteria 
such as target detection probability under a false alarm constraint, parameter estimation CRB or tracking mean square error (MSE)  are optimized while providing a desired data rate or quality of service (QoS) for communications users \cite{bicataes18}. In communications-centric designs,  the performance of communications, for example the sum rate is maximized under constraints on minimum tolerable sensing task performance. In multiantenna systems one can also optimize precoders, decoders, beampatterns or beam allocations, see \cite{zheng2019radar,BF_FD_JCAS_TCOM_2022}.  
Multiobjective or Pareto optimization methods consider communications and sensing objectives jointly and trade off among potentially conflicting objectives, see \cite{Fan_ISAC_6G_JSAC_2022}. There may be multiple Pareto-optimal solutions. In all the above ISAC optimization formulations, practical operational constraints are imposed such as total power or per antenna power, bandwidth, constant modulus or low PAPR, desired AF shape for sensing, and minimum SINR levels tolerated by cooperative users. 

As an example, the waveform optimization problem for Neyman-Pearson detection maximizes detection probability under a strict false alarm constraint in radar while ensuring communications users obtain a desired rate, and the total power budget is not exceeded may be formulated. 
The constraints on the desired rate in communications system are then expressed via {mutual information (MI)} in the form $\log(1+\text{SINR})$, {or in terms of tolerable outage probability} and allowed total transmitted power. The bandwidth $B$ that appears in capacity expressions for each subcarrier is assumed to be the same in a particular MC system.  The solution contains subcarrier selections, their power allocations and threshold selection for the detector. 

\subsection{ISAC Optimization using Machine Learning}

We will consider supervised learning and reinforcement learning (RL) methods for data-driven ISAC optimization. 

\subsubsection{{Optimization using Supervised Machine Learning} }Supervised machine learning methods rely on the availability of very large amounts of representative and labeled training data while requiring very few or no model assumptions. 
{Supervised learning can be applied to receiver learning, replacing standard methods with deep neural networks, or by parameterizing components of existing methods. This can make such methods more robust or efficient. Supervised learning can also be applied to transmitter learning, as well as to 
 end-to-end (E2E) learning, which simultaneously optimizes transmit signals in time-frequency-space and sensing receiver algorithms, e.g.,  for dealing with hardware impairments such as mutual coupling and inter-element spacing perturbations \cite{end2end_ISAC_2022}.}
 Simultaneous data-driven optimization of the ISAC beamformer and the sensing receiver enables the achievement of ISAC trade-offs comparable to those achievable with model-based processing when perfect knowledge of impairments is available. Model-based techniques, such as learning a dictionary of array steering vectors or directly learning parameterized impairments 
  offer lower computational complexity, more favorable ISAC trade-offs and better generalization capabilities to unseen testing data compared to model-free learning methods \cite{end2end_ISAC_2022}. Additionally, the structure of frequency-domain steering vectors can be leveraged to combat impairments that primarily affect delay estimation, such as CFO \cite{siso_ofdm_deep_unfolding}. Nevertheless, supervised E2E learning does have certain limitations, as it requires differentiable channel models for E2E learning and relies on labeled data in supervised learning, which may pose challenges in practical operation.

\subsubsection{{Optimization using} Reinforcement Learning}

RL methods are sequential decision making methods that maximize expected cumulative rewards while learning from experiences.
In RL an agent interacts with its operational environment by taking different actions.  After taking an action, the agent observes the environment state and receives a scalar reward. The reward defines what is important to the agent. Observations and the state may be subject to noise and uncertainties.  
Given the state of the environment, the agent takes the next action. Sequential decision-making processes are commonly formulated as Markov Decision Processes (MDP) or Partially Observable MDPs.

Model-free RL (MFRL) methods learn an optimal policy $\pi$ or a value function to autonomously improve the performance of an ISAC system by trial and error. 
A value function describes the expected discounted sum of rewards starting from a state and obtained directly from the experiences. MFRL needs to take the action to find the reward in different states or state-action pairs whereas a Model-Based Reinforcement Learning (MBRL) method may be able to predict it.  MBRL methods in ISAC take advantage of rich structural information on man-made communications and radar systems and the physics of the radio spectrum while learning from data.  {MBRL incorporates memory in the system by learning a predictive model of the state of the environment and how it evolves. It also models how the states lead to rewards and consequently uses the model to plan the policy or value function, and select the best actions.}
In general, MBRL allows for learning from experiences faster since it does not have to undergo numerous trials and errors as MFRL does. 

\begin{figure}
    \centering
    \begin{minipage}{0.48\textwidth}
        \centering
         \includegraphics[width=0.99\linewidth]{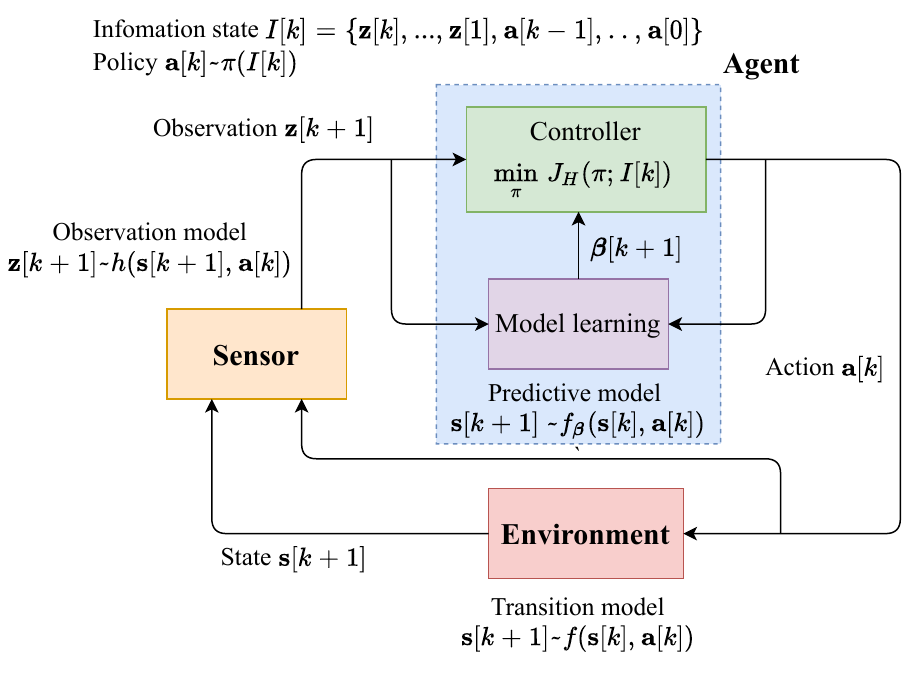}
    \caption{MBRL framework for allocating subcarriers and powers in MC ISAC. Transition model is learned and the controller chooses the actions using the learned model and MI-based rewards}
    \label{fig:lmpc}
    \end{minipage}
    \hspace{0.2cm}
    \begin{minipage}{0.48\textwidth}
        \centering
         \includegraphics[width=0.99\linewidth]{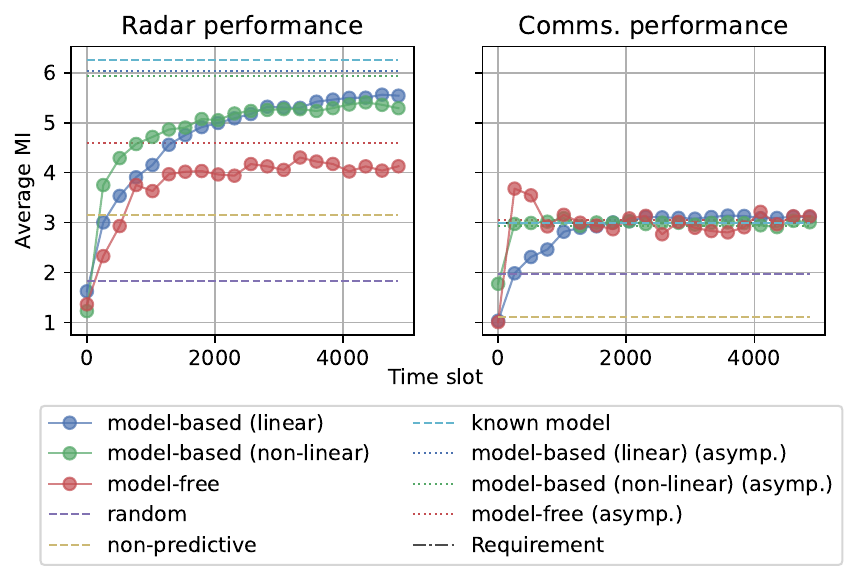}
    \caption{MBRL maximizes the sensing MI while ensuring the desired rates defined by MI for communications users. The results obtained using MFRL and linear MBRL are shown for comparison. MBRL adapts to changes faster}
    \label{fig:mi_comp}
    \end{minipage}
    \vspace{-0.22in}
\end{figure}

MBRL is related to control systems where an internal dynamic model is learned and the optimal control selected accordingly. As an example, we formulate the problem of subcarrier selection and power allocation in co-designed MC-ISAC system as a MDP with unknown transition model, see \cite{pulkkinen24} for details.
This model is 
depicted in Fig.~\ref{fig:lmpc}. 

The agent interacts with the environment state $\mathbf{s}[k]$ using actions ${\mathbf a}[k]$ where $k$ is the time index. The state transitions are described by a transition function $\mathbf{s}[k\!+\!1] \sim f_{\mathbf{\beta}}(\mathbf{s}[k], {\mathbf a}[k])$ which is a probability distribution and $\mathbf{\beta}$ are the model parameters to learn. The learned predictive model contains information on the radio environment. e.g. channel gains and experienced interferences at the receivers, and how they evolve when actions are taken. The agent observes the state using a receiver or feedback from cooperative users, giving the observation $\mathbf{z}[k\!+\!1] \sim h(\mathbf{s}[k\!+\!1], {\mathbf a}[k])$ where the observation model $h()$ is assumed to be known. 

The state {has a Markovian property and} can be predicted using a history of internal states and the transition function.  The internal state is comprised of the interference level and channel gain matrices 
$\left( \mathbf{\Psi}_s, \mathbf{\Psi}_u, \mathbf{R}_c, \mathbf{D}_s, \mathbf{D}_{u}, \mathbf{D}_{cs}, \mathbf{D}_{cu} \right)$ effective during time slot $k$. The subscripts $s,u$ refer to sensing and communications function, respectively, $\mathbf{\Psi}$ denote interference from non-cooperative users, $\mathbf{R}_c$ from cooperative users and $\mathbf{D}$'s are frequency domain channel matrices, with $\mathbf{D}_{cs}$ and $\mathbf{D}_{cu}$ denoting the interference channels from cooperative sensing and communications users, respectively.   The MI-based rewards are formed by using the state information about channels, noise plus interference powers, an consequently data rate via $\log(1+\text{SINR})$.  We distinguish between cooperative users that share awareness about the channels and interference, and other disturbances from non-cooperative or adversarial sources {causing  unintentional or intentional interference but do not share information}. 
The observation $\mathbf{z}$ is a noisy estimate of the internal state where $\mathbf{\Psi}_i$ is estimated from signal-free data, $\mathbf{R}_c$ is obtained via feedback or awareness sharing, and $\mathbf{D}_{i}$ and $\mathbf{D}_{ci}$ are estimated from known sensing waveforms or pilot sequences or using reciprocity. The model of the internal state is learned using logistic online regression in which the internal state components are quantized \cite{pulkkinen24}. 

The action is a vector ${\mathbf a} = [p_1, w_1,p_2, w_2, \dots, p_{N}, w_{N} ]$ that contains individual sub-carriers (or resource block of subcarriers) and their power allocations $p_n$ and selection variables $w_n \in \{s, u\}$ that indicate whether subcarrier $n$ is used for sensing ($w_n=s$) or communications ($w_n=u$).
There is a constraint on the maximum total power $\ptot$. In communications tasks, the MI reward is associated with data rate,. In sensing MI definition depends on the task, for example MI may be defined between the received radar echo and the target scattering coefficient (target impulse response). There is also an information theoretic relation between MI and MSE that could be used in tracking or parameter estimation.

A policy $\pi(I_k)$ is employed in RL that optimizes actions 
\begin{subequations}%\label{eq:mpc}
\begin{gather}
    \arg \max_\pi\; \mathbb{E}_{\pi} \left[ {\sum_{i=k}^{k+H-1} r(\mathbf{s}[i], \mathbf{s}[i\!+\!1], \mathbf{a}[i])| I_k} \right] \\ %\label{eq:mpc_reward} \\
    \text{subject to} \;  \mathbf{s}[i\!+\!1] \sim f_{\mathbf{\beta}}(\mathbf{s}[i], \mathbf{a}[i]),%\label{eq:mpc_transition} 
    \; \text{and} \; %\\
    \E{c_j(\mathbf{a}[i], \mathbf{s}[i])} \in \mathcal{C}_j,  \; \forall \; j=1,\dots,N_{\mathcal{C}}, %\label{eq:mpc_constraints}
\end{gather}
\end{subequations}
where $r(\mathbf{s}[k],\mathbf{s}[k+1],\mathbf{a}[k])$ is an immediate reward, $H$ is the planning horizon and $I_k$ is the history of actions and observations and the latter constraint contains all the operational constraints. In the above example, MI-based reward of the form  $\log(1+\text{SINR})$ is defined as $M_i[k\!+\!1] = \sum_{n=1}^{N} \mathbb{I}_{\{w_n[k]=i\}} \log\left(1 + \cqin[k\!+\!1] p_n[k] \right)$, where  $\cqin[k+1]$ is the SINR term (channel gain divided by the interference plus noise power) at the slot $k+1$, $\textit{Rx}$ $i$ and sub-carrier $n$. The minimum communications rate constraint is $ \E{M_u[k+1] | I_k} \geq C[k]$. A constraint on the interference power caused to cooperative users is imposed so that they achieve a tolerable SINR level \cite{pulkkinen24}.
%$\herm{\mathbf{D}_}{cs}[k+1]} \mathbf{D}_{cs}[k+1] \mathbf{p}[k] \leq \mathbf{c}[k]$,.
In terms of the MIs and the learned model in ISAC, the subcarrier and power allocation learning writes 
\begin{subequations}%\label{eq:mpc_obj}
	\begin{gather}
		\arg \max_{\mathbf{p}[k], \mathbf{w}[k] \in \mathcal{A}} \;  \E{M_s[k+1] | I_k} \\ %\label{eq:exp_mi}\\
		\text{subject to} \; \E{M_u[k+1] | I_k} \geq C[k],\label{eq:rate_const} \text{and} \; %\\
		\mathbf{D}_{cs}[k+1] \herm{\mathbf{D}}_{cs}[k+1]\mathbf{p}[k] \leq \mathbf{c}[k], %\label{eq:coop_const}
	\end{gather}
\end{subequations}
where the first constraint ensures desired data rate for communications users and the latter  $\mathbf{D}_{cs}[k+1] \herm{\mathbf{D}}_{cs}[k+1]\mathbf{p}[k] \leq \mathbf{c}[k]$ constrains the maximum interference power induced to cooperative users.
See \cite{pulkkinen24,bicataes18} for details on the learning method. To illustrate the performance of learning, the obtained MI levels are shown both for sensing and communication tasks in Fig.~\ref{fig:mi_comp}.
The achieved regrets are decidedly sublinear in all the different interference scenarios considered.

%%%%%%%%%%%%%%%%%%%%%%%%%%%%%%%%%%%%%%%%%%%%%%%%%%%%%%%

\section{Conclusions and Future Directions}

Most current and emerging wireless communication and broadcast systems, including 4G, 5G and future 6G and WLAN/WiFi employ multicarrier waveforms. MC signals are widely used for radars as well. They possess desirable properties both for broadband communications and RF sensing. Hence, they provide a strong candidate for ISAC in the emerging 6G systems. Furthermore, ISAC systems will benefit from ongoing convergence of communications and radar technologies including RF circuit and multifunction HW design, large aperture antenna arrays, fully adaptive transceivers and the ability to operate in a shared, densely used spectrum. MC and multiantenna ISAC provide tremendous research opportunities in theory, methods and implementation of emerging 6G systems. In this paper, we provided an overview of MC waveforms and related signal processing methods for ISAC systems. Both legacy OFDM and emerging MC modulations such as OTFS were considered. Highly relevant practical examples of MC and multiantenna (MIMO) transceiver signal processing in ISAC, and methods for mitigating or exploiting a variety of transceiver nonidealities were presented. The paper considered also the optimization of ISAC system performance for mutual benefit using classical optimization, data-driven and model-based machine learning methods. Emerging research topics include the use of Reflective Intelligent Surfaces (RIS) in multi-user, multi-target and multi-operator environments, THz technologies, security and privacy, machine learning and new applications and use cases for ISAC in 6G.

\section{Ackowledgements}
This work is supported, in part, by the SNS JU project 6G-DISAC under the EU's Horizon Europe research and innovation program under Grant Agreement No 101139130, by the Vinnova RADCOM2 project under Grant 2021-02568, by the Swedish Research Council (VR grant 2022-03007), by Business Finland under the project 6G-ISAC, by the Research Council of Finland under the grants \#345654, \#352754, \#357730, and \#359095, \#359094, EU INSTINCT and SAAB Finland CR, and by the National Science Foundation under grant no. 2433782 and the support in part by funds from federal agency and industry partners as specified in the Resilient \& Intelligent NextG Systems (RINGS) program.

%\balance
\linespread{1.25}
\bibliographystyle{IEEEtran}
\bibliography{references}

\end{document}